\documentclass[12pt,draftcls,journal,leterpaper,twosides,onecolumn]{IEEEtran}

\usepackage{cite}
\usepackage{amsmath,amssymb,amsfonts}
\usepackage{algorithmic}
\usepackage{graphicx,color}
\usepackage{textcomp}
\usepackage{comment}
\usepackage{xcolor}
\usepackage{tikz}
\usepackage{booktabs}
\usepackage{algorithm,algorithmic}
\def\BibTeX{{\rm B\kern-.05em{\sc i\kern-.025em b}\kern-.08em
    T\kern-.1667em\lower.7ex\hbox{E}\kern-.125emX}}
\AtBeginDocument{\definecolor{tmlcncolor}{cmyk}{0.93,0.59,0.15,0.02}\definecolor{NavyBlue}{RGB}{0,86,125}}

\usepackage{amsmath}

\DeclareMathOperator*{\argmin}{arg\,min}
\usepackage{bm}
\usepackage{graphicx,psfrag}
\usepackage{pgf}
\usepackage{amsmath,amssymb,amsthm,mathtools}
\usepackage{algorithm}
\usepackage{algorithmic}

\usepackage{amsmath}
\usepackage{graphicx}
\usepackage[margin=1in]{geometry}
\usepackage{auto-pst-pdf}
\usepackage{tabularx}
\usepackage{array}
\usepackage{colortbl}
\usepackage{xcolor}
\definecolor{verylightblue}{RGB}{224,240,255}
\usepackage{pifont} 
\usepackage{xcolor} 
\definecolor{customgreen}{RGB}{0, 160, 0} 
\definecolor{customred}{RGB}{160, 0, 0} 
\newcommand{\cmark}{\textcolor{customgreen}{\ding{51}}} 
\newcommand{\xmark}{\textcolor{customred}{\ding{55}}} 
\usepackage{glossaries}
\newacronym{fec}{FEC}{forward error correction}
\newacronym{arq}{ARQ}{automatic repeat request}
\glsunset{arq}
\newacronym{awgn}{AWGN}{additive white Gaussian noise}
\newacronym{qam}{QAM}{quadrature amplitude modulation}
\newacronym{qos}{QoS}{quality of service}
\newacronym{qoe}{QoE}{quality of experience}
\newacronym{iid}{i.i.d.}{independent and identically distributed}
\newacronym{snr}{SNR}{signal-to-noise ratio}
\newacronym{bcr}{BCR}{bandwidth compression ratio}
\newacronym{psnr}{PSNR}{peak signal-to-noise ratio}
\newacronym{mse}{MSE}{mean-squared error}
\newacronym{dl}{DL}{deep learning}
\newacronym{dnn}{DNN}{deep neural network}
\newacronym{deepjscc}{DeepJSCC}{deep joint source-channel coding}
\newacronym{jscc}{JSCC}{joint source-channel coding}
\newacronym{bsc}{BSC}{binary symmetric channel}
\newacronym{vimco}{VIMCO}{variational inference for Monte Carlo objectives}
\newacronym{ml}{ML}{maximum likelihood}
\newacronym{gdn}{GDN}{generalized divisive normalization}
\newacronym{ber}{BER}{bit error rate}
\newacronym{bler}{BLER}{block error rate}
\newacronym{ldpc}{LDPC}{low density parity check}
\newacronym{uep}{UEP}{unequal error protection}
\newacronym{csi}{CSI}{channel state information}
\newacronym{relax}{RLACS}{rateless and lossy autoencoder channel and source}
\glsunset{relax}
\newacronym{crc}{CRC}{cyclic redundancy check}
\newacronym{5g}{5G}{5th generation}
\newacronym{6g}{6G}{6th generation}
\newacronym{sla}{SLA}{service level agreement}
\newacronym{tcp}{TCP}{transmission control protocol}
\newacronym{udplite}{UDP-Lite}{user datagram protocol--lightweight}
\newacronym{ai}{AI}{artificial intelligence}
\newacronym{per}{PER}{packet error rate}
\newacronym{harq}{HARQ}{hybrid automatic repeat request}
\glsunset{harq}
\newacronym{ae}{AE}{autoencoder}
\newacronym{pca}{PCA}{principle component analysis}

\usepackage{epstopdf}

\usepackage{pgfplots}
\usepackage{pgfplotstable}
\pgfplotsset{compat=1.17}
\usepackage[caption=false,font=footnotesize]{subfig}

\newcommand{\secref}[1]{Sec.~\ref{#1}}

\newcommand{\textsizescale}{0.55} 
\newcommand{\textsizescaleM}{0.75} 
\newcommand{\imagescalesize}{0.6} 
\newcommand{\LabelFontSize}{0.6} 
\newcommand{\archfigscale}{0.42} 

\newtheorem{definition}{Definition}

\newtheorem{remark}{Remark}

\usepackage{mathptmx}
\usepackage[mathscr]{eucal}

\usepackage{amsthm}
\usepackage{amsfonts}
\usepackage{amssymb}
\usepackage{amsmath}
\usepackage{mathtools}
\DeclareMathAlphabet{\mathcal}{OMS}{cmsy}{m}{n}

\DeclareSymbolFont{Letters}{OML}{cmm}{m}{it}
\DeclareMathSymbol{\alpha}{\mathalpha}{Letters}{11}
\DeclareMathSymbol{\beta}{\mathalpha}{Letters}{12}
\DeclareMathSymbol{\lambda}{\mathalpha}{Letters}{21}
\DeclareMathSymbol{\Lambda}{\mathalpha}{Letters}{3}
\DeclareMathSymbol{\pi}{\mathalpha}{Letters}{25}
\DeclareMathSymbol{\rho}{\mathalpha}{Letters}{26}
\DeclareMathSymbol{\sigma}{\mathalpha}{Letters}{27}

\makeatletter
\def\@IEEEsectpunct{:\ \,}
\def\paragraph{\@startsection{paragraph}{4}{\z@}{1.5ex plus 1.5ex minus 0.5ex}%
{0ex}{\normalfont\normalsize\bfseries}}
\makeatother

\usetikzlibrary{calc}
\usetikzlibrary{decorations.pathreplacing,decorations.markings,shapes.geometric}
\tikzset{naming/.style={align=center,font=\small}}
\tikzset{antenna/.style={insert path={-- coordinate (ant#1) ++(0,0.25) -- +(135:0.25) + (0,0) -- +(45:0.25)}}}
\tikzset{station/.style={naming,draw,shape=dart,shape border rotate=90, minimum width=10mm, minimum height=10mm,outer sep=0pt,inner sep=3pt}}
\tikzset{mobile/.style={naming,draw,shape=rectangle,minimum width=12mm,minimum height=6mm, outer sep=0pt,inner sep=3pt}}
\tikzset{radiation/.style={{decorate,decoration={expanding waves,angle=90,segment length=4pt}}}}

\tikzset{radiation/.style={{decorate,decoration={expanding waves,angle=90,segment length=4pt}}},
         BS/.pic={
        code={\tikzset{scale=5/10}
            \draw[semithick] (0,0) -- (1,4);
            \draw[semithick] (3,0) -- (2,4);
            \draw[semithick] (0,0) arc (180:0:1.5 and -0.5) node[above, midway]{#1};
            \node[inner sep=4pt] (circ) at (1.5,5.5) {};
            \draw[semithick] (1.5,5.5) circle(8pt);
            \draw[semithick] (1.5,5.5cm-8pt) -- (1.5,4);
            \draw[semithick] (1.5,4) ellipse (0.5 and 0.166);
            \draw[semithick,radiation,decoration={angle=50}] (1.5cm+8pt,5.5) -- +(0:2);
            \draw[semithick,radiation,decoration={angle=50}] (1.5cm-8pt,5.5) -- +(180:2);
  }}
}

\tikzset{radiation/.style={{decorate,decoration={expanding waves,angle=90,segment length=4pt}}},
         node/.pic={
        code={\tikzset{scale=5/10}
            \filldraw[semithick] (0,0) circle(10pt);
            \draw[semithick,radiation,decoration={angle=45}] (0,12pt) -- +(90:2);
  }}
}

\newcolumntype{C}{>{\(\displaystyle}c<{\)}@{}} 
\newcolumntype{L}{>{\(\displaystyle}l<{\)}@{}} 
\newcolumntype{R}{>{\(\displaystyle}r<{\)}@{}}

\newcommand{\figref}[1]{Fig.~\ref{#1}}

\newcommand{\tabref}[1]{Table~\ref{#1}}
\newcommand{\Algref}[1]{Alg.~\ref{#1}}

\newcommand{\remref}[1]{Remark~\ref{#1}}
\newcommand{\defref}[1]{Definition~\ref{#1}}

\usepackage{hyperref}

\begin{document}
\bstctlcite{MyBSTcontrol}

\title{\parbox{\textwidth}{\centering 
\fontsize{19.5pt}{24pt}\selectfont Rateless Joint Source-Channel Coding, and a \\ Blueprint for 6G Semantic Communications System Design}}
\author{Saeed R. Khosravirad \IEEEauthorblockA{}

\IEEEauthorblockA{Nokia Bell Labs}\\
e-mail: saeed.khosravirad@nokia-bell-labs.com
}

\maketitle

\begin{abstract}
    This paper introduces rateless joint source-channel coding (rateless JSCC). The code is \emph{rateless} in that it is designed and optimized for a continuum of coding rates such that it achieves a desired distortion for any rate in that continuum. We further introduce rate-adaptive and stable communication link operation to accommodate rateless JSCCs. The link operation resembles a “bit pipe” that is identified by its rate in bits per frame, and, by the rate of bits that are flipped in each frame. Thus, the link operation is \emph{rate-adaptive} such that it punctures the rateless JSCC codeword to adapt its length (and coding rate) to the underlying channel capacity, and is \emph{stable} in maintaining the bit flipping ratio across time frames. 
    
    Next, a new family of autoencoder rateless JSCC codes are introduced. The  code family is dubbed RLACS code (read as \emph{relax code}, standing for ratelss and lossy autoencoder channel and source code). The code is tested for reconstruction loss of image signals and demonstrates powerful performance that is resilient to variation of channel quality. RLACS code is readily applicable to the case of \emph{semantic} distortion suited to variety of semantic and effectiveness communications use cases. 
    
    In the second part of the paper, we dive into the practical concerns around semantic communication and provide a blueprint for semantic networking system design relying on updating the existing network systems with some essential modifications. We further outline a comprehensive list of open research problems and development challenges towards a practical 6G communications system design that enables semantic networking.
\end{abstract}

\section{Introduction}
\label{sec:intro}

The concepts of \emph{semantic} and \emph{effectiveness} communication were raised by W. Weaver in a preface to Shannon's mathematical theory of communication---while referring to Shannon's work as a solution to  \emph{technical} communication problem---as what should come next beyond the technical communication \cite{shannon1998mathematical}. Several attempts are made to formalize those concepts (see for instance \cite{floridi2004outline,isaac2019semantics,zhong2017theory,shao2024theory,saz2024model}), yet a comprehensive mathematical formalization and a framework to evaluate the performance limits of semantic communications system remain  open problems. 

Specifically, a formal definition of the semantic  problem that differentiates it against the technical  problem  towards a meaningfully different communication networking solution, is  not available. The notion of ``conveying  the desired meaning'', as opposed to ``accurate reconstruction of bits/symbols'', was alluded to by Weaver to differentiate semantic against technical problems. The former is thus seen by the literature mostly  as a source coding problem with majority effort focused on lossy \gls{jscc}, but the  impact on what we call \emph{communication network} is yet unclear. In source coding, the differences are  evident and semantic compression has already provided meaningful engineering solutions: for instance, the hierarchical codecs used for image \cite{li2022deep,huang2021deep,yang2015visual,7226830} and video \cite{zhu2003hierarchical,zhai2005joint} signals can distinguish between semantic vectors and perceptual elements in the signal and compress them at unequal rates according to their importance in reconstruction loss. Generative models have been successful in using  auto-regressive or diffusion-based decoders to expand on the resolution and perceptual quality of a digital signal based on  understanding of the generative priors \cite{kingma2022autoencodingvariationalbayes,chen2024generative} and the semantic vectors \cite{tang2024evolving} of a given data distribution. In language processing, compressing word and sentence tokens to \emph{embeddings} is an example of semantic compression---in fact, in the embedding space one can subtract   the embedding vector of word tokens and convert the difference into a word token, while the whole operation has \emph{meaning} in the word space too, e.g., $   vector (\text{``king''}) - vector (\text{``man''}) +  vector (\text{``woman''}) \Rightarrow vector (\text{``queen''})$  \cite{mikolov2013efficient}. 

\subsection{The Issue with Current Communication Networking Paradigm}

Semantic communication problem therefore mostly appears to have to do with  source coding and less with  communication networking, however, essential modifications to the way network treats the  messages can improve the  semantic extraction and reconstruction efficiency. The required changes stem from the intense focus of the current communication networking paradigm on  \emph{error-free} and \emph{packet-ized} communication. 
Within the network,  packets go through a sophisticated and well-optimized process of routing, error correction coding, channel symbol mapping, retransmissions, etc. By the end of their finite latency budget, the packets are only passed to the destination if successfully reconstructed---i.e., error-free. Otherwise, there will be a packet erasure. The \emph{cliff effect} and \emph{leveling-off effect} associated to separate source and channel coding  in \cite{gunduz2024joint} is mainly due to this. If a packet with small damage is still passed to the application, one that deploys error-tolerant source compression (or, \gls{jscc}), then the performance will smoothly and more gracefully degrade with channel quality decreasing (we  explore examples of such effect in this paper).

The current networking paradigm is inherently optimized for error-free (but, erasure-tolerant) communication. As a result, applications that tolerate loss (or, distortion) in message reconstruction are potentially not getting the best they can out of network's capacity. For instance, a common trait for video application codecs is to use hierarchical (or layered) compression of a frame into several unequally important data packets, while the network treats that unequal importance at \gls{qos} level. Network is often unaware of the relation among those several packets\footnote{In more advanced networking solutions, application-awareness can help the network understand such relations and make more cohesive decisions in this sense.}. 
From application's perspective, each  packet goes through an erasure channel with different reliability---it is either delivered error-free or not delivered within the  latency budget. Application, then uses  \emph{error concealment} techniques \cite{aign1995temporal} to recover  the lost packets. Overall, it is unclear whether such multi-layered  process is  near optimal in utilizing capacity (and, efficient), while there are concerns with the level of overhead imposed by such operation on both the network and the application \cite{lan2021semantic}. 

\subsection{Joint vs Separate Source-Channel Coding}

Optimality of separate source and channel coding  solutions \cite{shannon1959coding} hold only in the limit of infinite blocklength.  The assumptions that are needed for optimality frequently  don't  hold in practice  where there exist constraints on end-to-end latency and implementation complexity  \cite{zhai2005joint}. In lossy communication and finite blocklength regeime, \cite{6408177} proves sub-optimality of separation.  Joint solutions outperforming separate desgins have presistantly reported in the literature, e.g., see examples in \cite{10747747,lan2021semantic}. On the other hand,  \emph{separation}-based solutions have become indispensable to telecommunication systems,  offering ease of interoperability, modular and independent development of devices and network systems, and simplicity of  design. But another important benefit of separate design is channel agnosticism offered to the source compression, allowing the channel code to adapt to the channel state when needed.

In fact, there appears to be a critical challenge related to semantic (or, let's say, \emph{lossy}) communication of information, and that is the lack of channel awareness at the applications. 
The semantic application has the uncoded source information---exchanging that with the network breaches privacy of the source and requires complex interface between application and the network, thus source coding must happen at the application. Meanwhile, the network has the channel state information---exchanging that with the application exacerbates aging effect of \gls{csi},  increases operation overhead, and in complex networking systems is virtually impossible to be routed back to the application. As a result, in time-varying channel state conditions, the otherwise ideal \gls{jscc} solutions can  readily become suboptimal because ``channel'' is unknown. This appears to be the chief practical challenge for realizing the performance gain of \gls{jscc} against separate design. In practical networks the challenge is  compounded---it is not only the wireless channel quality that shapes the \emph{state} of the link, but also the randomness of the load on the network resulting from multi-user nature of load and heterogeneity   of users, and the multi-hop nature of majority of communication links (see discussion in \secref{sec:sketch})---that makes central tracking of the \emph{state} of the link impractical.






\begin{figure*}[ht]
\begin{center}
\psfrag{ss}[c][c][1]{$\mathbf{s}$}
\psfrag{f}[c][c][1]{$f$}
\psfrag{uu}[c][c][1]{$\mathbf{u}$}
\psfrag{ll8}[c][c][1]{$r$}
\psfrag{xx}[c][c][1]{$\mathbf{x}$}
\psfrag{ch}[c][c][1]{$\mathsf{P}(\mathbf{y}|\mathbf{x})$}
\psfrag{yy}[c][c][1]{$\mathbf{y}$}
\psfrag{ll9}[c][c][1]{$r^{*}$}
\psfrag{uh}[c][c][1]{$\hat{\mathbf{u}}$}
\psfrag{gg}[c][c][1]{$g$}
\psfrag{sh}[c][c][1]{$\hat{\mathbf{s}}$}
\psfrag{lll}[c][c][1]{application source}
\psfrag{nn}[c][c][1]{network}
\psfrag{rr}[c][c][1]{application destination}
\psfrag{ll0}[c][c][1]{(source) $N$}
\psfrag{ll1}[r][r][1]{$K$}
\psfrag{ll2}[r][r][1]{$K-L$}
\psfrag{ll3}[r][r][1]{$K-L$}
\psfrag{ll4}[r][r][1]{$K-L$}
\psfrag{Ll5}[r][l][1]{$L$}
\psfrag{ll7}[r][r][1]{(null)}
\psfrag{ll6}[c][c][1]{(recovered) $N$}
\includegraphics[width=.85\textwidth,keepaspectratio]{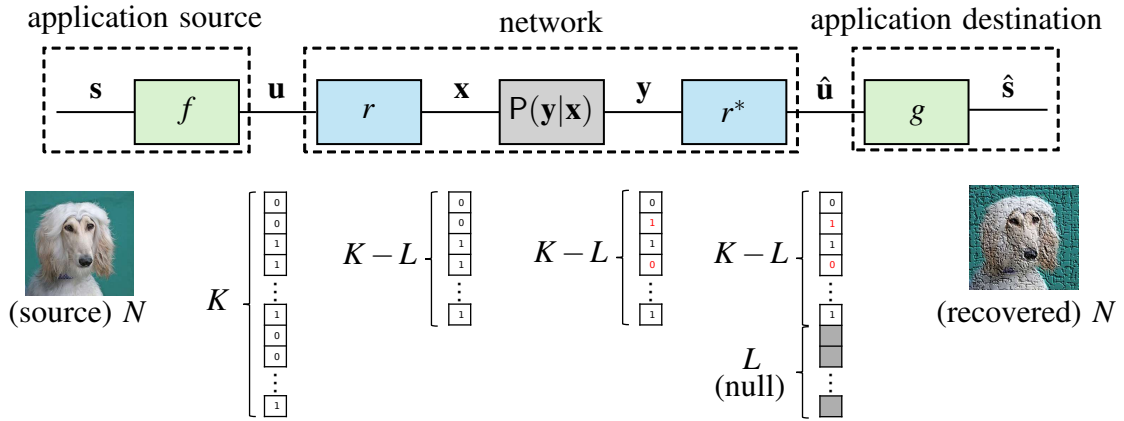}
\caption{Rateless JSCC and rate-adaptive link for semantic communication. At the bottom, $\mathbf{x}$ and $\mathbf{y}$ are shown as binary sequences, as an example. In general, the symbols selected for communication are chosen from a constellation based on the estimated channel knowledge at the transmitter. The bits in red represent the flipped bits due to communication error, while the shaded blocks represent the null bits at the decoder caused by puncturing in the network.}
\label{fig:system-model}
\end{center}
\end{figure*}

\subsection{Motivation}


The proposed rateless \gls{jscc} coding framework in this work  aims to resolve this very issue by hosting the \gls{jscc} functionality at the application while allowing the network to adjust the coding rate based on the knowledge of \gls{csi}. Particularly, the responsibility of the network is simplified to stabilizing the reliability of the end-to-end link (measured in ratio of flipped bits in each frame) and then adapting the coding rate to \gls{csi} by \emph{puncturing} the binary codeword received from the application. This relaxes the need for continuous coordination between the application and the network, and forms a cooperative game-theoretic venture between the network and the application. For this to function properly, the \gls{jscc} code at the application must perform well for a range of coding rates that are among network's possible choices and maintain optimality in the \emph{rate-distortion} tradeoff over that range.

Conventional source codes recognize rate-distortion tradeoff most commonly through quantization control. Rate control is therefore possible for those codes if performed by the source. However, commissioning the network with  rate control using the conventional source codes results in sub-optimal \gls{jscc} performance.

The proposal in this work thus has two main elements: a rateless \gls{jscc} coding framework for the application, and a rate-adaptive and stable end-to-end link operation framework for the network. In the following (as illustrated in  \figref{fig:system-model}),  \emph{application} refers to the user (a human and/or a machine running an application), and,  \emph{network}  refers to the service provider that establishes an end-to-end link for the application (possibly consisting of multitude of wireless and wired hops, switching and processing nodes). We also use the term \emph{frame} to refer to a segment of data bits that is communicated at a time by the application---the term is to be differentiated with \emph{packet} in technical communication which implies integrity requirement (see further discussion on this in \secref{sec:blueprint}).

\subsection{Prior Work}
\label{sec:priorwork}

The literature on semantic communication and lossy \gls{jscc} is rich, especially the research on leveraging the power of \glspl{dnn} in realizing semantic communication. The overview papers on the topic that raise interesting open research problems are outlined later in \secref{sec:openresearch}. Here, we study the literature around a few  points that are the most relevant to our rateless \gls{jscc} proposal.

\subsubsection{Hierarchical Compression and Variable-Length Coding}

The proposed rateless \gls{jscc} codes shouldn't be mistaken by the so-called variable-length \gls{jscc} schemes, such as in~\cite{10198383,8445924}, where the code length is varied according to network-provided rate or source word length. In contrast, a rateless \gls{jscc} code generates a codeword that can cover a continuum of compression rates  without knowing which rate will be used by the network. Thus, the aim of the code design is to maintain a rate-distortion tradeoff for all  the supported rates. 

Use of rateless codes such as Raptor codes together with hierarchical source compression was proposed in \cite{bursalioglu2008lossy,bursalioglu2013joint} to provide a continuum of coding rate that can be adapted to the channel capacity. The medium is first  quantized into multiple bitplanes where each bitplane is then channel coded using Raptor codes, allowing for the coding rate to be adapted to the transmission channel capacity. In effect, the hierarchical source quantization and the channel coding remain separate while each quantization  hierarchy requires a separate coding rate matching. In contrast, in this work we target codes that achieve  true joint source-channel coding effect with single coding rate matching, and propose an example of those codes using deep neural networks that can be flexibly made rateless according to the desired range of coding rates.

\subsubsection{Unequal Importance of JSCC Coded Bits}

In practical source coding  it can be observed that  the individual coded bits incur different levels of relative importance or sensitivity. As a result, channel errors in different bit positions can affect the reconstructed signal differently. Such effect is apparent specially in variable-length source coding~\cite{748887}, such as entropy coding schemes---known to be otherwise optimal over \emph{noiseless} channel.

Several approaches have acknowledged this effect in developing \gls{jscc} techniques  under noisy channel conditions. Some of those solutions focus on designing the source and channel code to limit propagation of channel errors, e.g., by improving  synchronization for variable-length source codes~\cite{selfsyncvariablejscc}, or, through packetization during channel coding~\cite{138998}. Other solutions focus on  specifically weighted channel coding protection for different coded bit positions~\cite{1095155,748707}. A more recent work capitalizes on further accented unequal importance of the coded bits and proposes to design \gls{ae} type channel codes to unequally protect the source encoded bits with non-uniform importance~\cite{tung2024multilevelreliabilityinterfacesemantic}. 

\subsubsection{Use of Deep Neural Networks for JSCC}

This is a developing and rich part of the literature. \cite{xie21_deep_learn_enabl_seman_commun_system}, \cite{jankowski21_wirel_image_retriev_edge} and
\cite{bourtsoulatze19_deep_joint_sourc_chann_codin} are  examples of studying deep learning-based  \gls{jscc}. The latter targets wireless image transmission and demonstrates improved performance over traditional separate source and channel coding schemes, particularly in low  \gls{snr}. \cite{xu2023deep} proposes a \gls{snr}-adaptive \gls{jscc} solution based on deep learning, where the \gls{jscc} is aware of the channel estimate and uses that to properly encode to the capacity of the channel. This implies exchange of channel state between the network and the application. \cite{kalkhoran23_secur_deep_jscc_again_multip_eaves}, among others,  address the issue of security in \gls{dnn} based \gls{jscc}. In another interesting proposal, \cite{tang2024evolving} studies the approach of enhancing semantic communication  efficiency over time by improving the learned semantic vectors of communicated images and adding them to the shared contexts between the two ends of the link.

\subsubsection{Rateless Compression}

\glspl{ae} have proven powerful in compression tasks, thanks to  nonlinear activation functions such as rectified linear unit (ReLU). Compared to majorly linear techniques such as \gls{pca}, \glspl{ae} can more efficiently extract   features (a.k.a., latent representations) of  the data. This shortcoming of \gls{pca}-based techniques was addressed in \cite{scholz2008nonlinear}, where  non-linear \gls{pca} and hierarchical \glspl{ae} were discussed. In essence, sorted principle components of a liner (or, non-linear) \gls{pca} can be dropped out (i.e., punctured), starting from the least significant order, to create a \emph{rateless} behavior in the compression process. However, this is not generally true for \glspl{ae} since the latent
variables are  learned to be \emph{uniformly} important. Similar shortcoming is persistent  in sparse \glspl{ae} too \cite{ng2011sparse}, where the \gls{ae} learns  latent representations  that can be randomly dropped out, thus providing  rate-control mechanism for compression. This was investigated by \cite{koikeakino2020stochasticbottleneckratelessautoencoder} in developing a special case of \glspl{ae}, referred to as rateless \gls{ae}. 

The rateless \gls{ae}  in \cite{koikeakino2020stochasticbottleneckratelessautoencoder} is based on \emph{TailDrop} regularization which is stochastic in nature, but unlike sparse \glspl{ae}, imposes  a non-uniform dropout rate to the latent space. Specifically, the dropout rate in TailDrop monotonically increases   from head neurons to tail neurons. The result is an autoencoder with a stochastic bottleneck and rateless behavior that is
adaptable to a continuum of compression rates. The rateless \gls{ae} structure was utilized later by \cite{10437920} to construct a rateless deep \gls{jscc} solution for Point Cloud signals. 
The rateless \gls{jscc} in \cite{10437920}  allows the  code rate to be adapted, however, it doesn't assume a quantized (binary) interfacing between network and application entities and appears to unify the function of those entities. In practice,  training \gls{ae}  with binary latent space while applying a stochastic dropout, as was proposed in \cite{koikeakino2020stochasticbottleneckratelessautoencoder}, proves  highly complex---the training must converge while experiencing randomness of the dropout regularization, randomness of bit flipping due to channel errors, and the quantization of the latent space into binary. Therefore, in the proposed solution in this paper we avoid the random dropout of the latent space to improve  convergence of  \gls{ae} training, and instead, propose a new sequential training mechanism that makes the code  rateless and imposes a  non-uniform and monotonically decreasing importance of the coded latent features.

\subsection{Contributions}

The main contributions of this paper can be summarized as follows:

\begin{enumerate}
    \item We introduce the new framework of  rateless \gls{jscc}. Rateless JSCC is unaware of the state of the network (including channel state and load variation) but anticipates  time-variation in the state, hence optimizes the code to be rate-compatible and covers a continuum of such variations\footnote{The term ``rateless'' is  borrowed from the rateless channel codes \cite{6145513} but has a different  effect here. That is, rateless JSCC encodes and transmits to the highest  rate, but the decoder receives  a portion of the codeword (i.e., the portion allowed by the network) at a lowered rate. Regardless of the rate, the code must try to minimize the distortion.}. Further, we introduce a new rate-adaptive and stable communication link approach. The proposed link operation turns the underlying channel into a \emph{bit pipe} that is identified by the number of bits it transfers in a single time frame and the rate of bits that are flipped during the transfer. The link accommodates rateless JSCC by being \emph{rate-adaptive}, such that it punctures the codeword  to adapt its length (and coding rate) to the underlying channel capacity, and is  \emph{stable} in maintaining the bit flipping ratio across time frames. The puncturing follows an order of bits that is known by both the network and the application, e.g., from the end of the codeword, such that the more important information is encoded into bits that are less likely to be punctured. Due to the puncturing, the link introduces a new paradigm of communication networking where the network is allowed to throw away part of the data based on pre-agreed terms with the application. 
    \item We propose an \gls{ae}  solution to  rateless JSCC. The proposed code is dubbed rateless and lossy autoencoder channel and source (RLACS) code (read as relax code!) and is built using residual attention network architecture \cite{wang2017residual} with binary latent space. The code is trained and tested on image signals considering reconstruction loss as the objective distortion and demonstrates powerful performance against channel state variation thanks to the rateless JSCC structure.
    \item A deep dive into practical implications of a semantic communication system is presented and a blueprint of a future end-to-end communication system design is provided. The proposed system design is aimed at enabling semantic and effectiveness communication with some essential modifications to the existing network systems, while maintaining interoperability and coexistence with other application types. The discussion is complemented with a comprehensive list of opern research and development problems towards a \gls{6g} semantic communications system.
\end{enumerate}

\subsection{Organization of the Paper} 

\secref{sec:prelim} describes the problem of semantic communication over noisy and faded channel; \secref{sec:proposal} introduces rateless JSCC for application operation and rate-adaptive and stable link for network operation; \secref{sec:relax} presents an \gls{ae} solution to rateless JSCC based on sequential training of a multi-DNN attention residual network architecture, and provides  numerical experiment results for the proposes RLACS code; \secref{sec:blueprint} provides a blueprint for next generation communication networks with semantic communication capabilities and outlines a comprehensive list of open research problems towards 6G semantic communication system design; and, finally, \secref{sec:conclusions} concludes the paper.

\section{Preliminaries and Problem Statement}
\label{sec:prelim}

We study the problem of semantic communication over a  channel with time-varying gain $\sqrt{\gamma_t}$ and \gls{awgn} $\mathbf{n}$. The channel is assumed to be block fading, that is, it stays constant over a  blocklength  $\bar{W}$ (in Hz-s), after which, it changes to a new independent channel, where $t$ is the block index\footnote{Block fading assumption is taken for simplicity. The definitions and the proposed solution in the paper apply  to frequency selective fading and doubly selective fading channel.}. 

A \emph{network operator} (or, network, for simplicity) operates over the channel to create a semantic communication link between  \emph{application source} and  \emph{application destination} nodes (see \figref{fig:system-model}). The network has knowledge of the \gls{csi} $\tilde{\gamma_t}$, which is an estimate of the channel gain ${\gamma_t}$ (i.e., it is either delayed or noisy, or both). Hereafter, the term \emph{block} refers to a coherence block of the underlying channel, while \emph{frame} refers to a segment of data communicated by the application over a block. 

The application uses a \gls{jscc} code   with   encoder $f$ and decoder $g$ (defined in the following section). 
During  block $t$, the source encodes an input medium $\mathbf{s_t}$ (e.g., an  image signal) into a binary sequence \( \mathbf{u_t} = f(\mathbf{s_t}) \) of length $K$ and delivers it to the network. Network's job is to deliver a binary sequence $\hat{\mathbf{u}_t}$ to the application destination, where the sequence is decoded into \( \hat{\mathbf{s}_t} = g(\hat{\mathbf{u}_t}) \). 

The network operator has a total transmit power limit of $P_t \leq \Bar{P}$ and a total channel use  limit of $W_t \leq \bar{W}$, during  block $t$. Both $P_t$ and $W_t$ depend on the resource scheduling strategy of the network among multitude of users. Meanwhile, $\tilde{\gamma_t}$ depends on the randomness of the underlying channel. Together, $(\tilde{\gamma_t},P_t,W_t)$ forms the time-varying \emph{state} of the network and the end-to-end link provided for the application. For simplicity, we assume that the transmit power and channel use limits are constant over blocks. The goal of this semantic communication system is to minimize the expected distortion $\mathsf{d}({\mathbf{s}_t}, \hat{\mathbf{s}}_t)$ for each block $t$.

\begin{remark}\label{rem:exchange}
The network and the application do not exchange the uncoded source information ${\mathbf{u}}$, nor the state of the network. Therefore, the encoder $f$ (and  the decoder $g$) are unaware of the state of the channel and the network is unaware of the distribution and instance of the source signal. 
\end{remark}

\begin{remark}
    The goal laid out above is distinct from that of the classical \gls{jscc} excess distortion criterion, i.e., $\mathsf{Pr}[\mathsf{d}({\mathbf{s}}, \hat{\mathbf{s}}) > d] \leq \delta$. The intention, from the application perspective, is to minimize the expected distortion given the randomness in the \emph{state} of the network, $(\tilde{\gamma_t},P_t,W_t)$. It is conceivable to further assume that the application is not aware of the distribution of those random variables either. Additionally, average distortion over several blocks is not a suitable goal for still media forms such as images. 
\end{remark}

In the following, the index $t$ is omitted for notation simplicity, unless specifically needed.

\begin{table}[!ht]
\caption{List of Notations}
\label{Tab:notations}
\centering
\footnotesize 
\begin{tabularx}{.95\columnwidth}{|>{\raggedright\arraybackslash}p{0.15\columnwidth}||>{\raggedright\arraybackslash}X|}
\hline 
\rowcolor{lightgray} \textbf{Notation} & \textbf{Description}  \\
\hline \hline
$N$; $K$   &  Dimension of the uncoded source signal;  dimension of the coded  signal. \\ 
\hline
$L$ ; $\bar{L}$   &  Random variable denoting bit puncturing size;  the maximum value $L$ takes. \\ 
\hline
$\sigma_o$; $\sigma_e$  &  \gls{awgn} spectral density; variance of the channel estimation noise. \\ 
\hline
$\mathsf{F}$; $\mathsf{P}$; $\mathsf{Pr}$   &  Cumulative distribution;  probability distribution;  probability of random events. \\ 
\hline
$\mathsf{d}$; $\mathbf{d}$   &  Distortion function; vector of distortion values. \\ 
\hline
$\mathbf{s}$; $\hat{\mathbf{s}}$   &  Uncoded source signal; its reconstruction. \\ 
\hline
$\mathbf{u}$; $ \hat{\mathbf{u}}$   &  Coded source signal; its reconstruction. \\ 
\hline
$R$; $R_{lo}$; $R_{hi}$    &  Coding rate; lowest coding rate; highest coding rate. \\ 
\hline
$M$; $c$    &  Modulation order; number of possible modulation orders. \\ 
\hline
$q$; $q_o$    &  Bit flipping ratio; target bit flipping ratio. \\ 
\hline
$P$; $\bar{P}$    &  Transmit power; maximum transmit power. \\ 
\hline
$W$; $\bar{W}$    &  Transmission channel use; maximum transmission channel use. \\ 
\hline
$f$; $g$    &  Encoder function; decoder function. \\ 
\hline
$C_i$; $F$    &  Input codeword size for decoder $i$; number of encoder-decoder pairs for RLACS code. \\ 
\hline
$\mathsf{Q}$; $\mathsf{Q}^{-1}$ & Mapping between \gls{snr}, modulation order $M$ and \gls{ber};  inverse of $\mathsf{Q}()$ with respect to SNR. \\ 
\hline
$\sqrt{\gamma}$; $ \tilde{\gamma}$ &  Channel gain; noisy estimate of $\gamma$. \\ 
\hline
$\theta$; $\phi$ &  DNN parameter set for encoder function; the decoder parameters. \\ 
\hline
$\varnothing$; $\bm{\varnothing}$ &  Null value; a vector of null values. \\ 
\hline
\end{tabularx}
\end{table}

\section{Proposed Solution: Rateless and Lossy Source-Channel Codes}
\label{sec:proposal}

With the limitation in information exchange described in \remref{rem:exchange}, we propose a joint venture between the network and the application where the following  set of ground rules are established initially between the two and followed during the course of semantic communication service. 
\begin{itemize}
    \item The network and the application exchange over a binary  interface. Thus, the end-to-end channel from the application's perspective is a vector binary channel. The network stabilizes the end-to-end  link by maintaining a bit flipping ratio (or \gls{ber}) across frames. 
    \item The source-channel code operated by the application is \emph{rateless} in that it is designed and optimized for a 
    range of rate values $[R_{lo}, R_{hi}]$. The source signal is  encoded at rate $R_{hi} = K/N$ and the network decides on the actual rate $(K-L)/N$ within that range, by puncturing $L$ bits out of a codeword it receives from the application. Without loss of generality, we assume that the $L$ bits are selected from the end of the codeword.
\end{itemize}

The puncturing length $L$ is modeled as a non-negative integer discrete random variable with distribution $\mathsf{P}_L$, where $L \in \{0, \ldots, \bar{L}\}$. At each time $t$, it is determined by the \emph{scheduler} at the network (denoted by function $r$ in \figref{fig:system-model}) based on the state of the network, $(\tilde{\gamma_t},P_t,W_t)$.

\begin{remark}
    The joint venture of the network and the application in this problem can be seen as a cooperative interaction in game theory. The players, network and application, exchange some ground rules---known as service level agreement or link configuration in telecommunication terminology---and cooperate towards achieving the goal. This also implies network's cooperative effort to maximize $P_t$ and $W_t$ within its practical limits.
\end{remark}


\subsection{Network's Operation: Rate-Adaptive and Stable Link for  Semantic Communications}
\label{sec:ratelessradiolink}

Let us focus on the single-user setting, where total transmit power $\bar{P}$ and total channel use $\bar{W}$ is allocated to the user. The block fading channel provides a \gls{snr} of $\bar{P}\gamma / (\sigma_n \bar{W})$, where $\sigma_n$ is the \gls{awgn} noise spectral density. The network uses modulated communication, e.g., from \gls{qam} constellation, to transmit the bits over  faded and noisy channel. Using the knowledge of  $\tilde{\gamma}$, the conditional cummulative distribution of $\gamma$ is denoted by $\mathsf{F}_{\gamma | \tilde{\gamma}}$.
The network can choose a  modulation constellation to guarantee a target bit-flipping ratio between $\mathbf{x}$ and $\mathbf{y}$. 

Let us assume the modulation  constellation order $M$ is chosen from a finite ordered set $ \{M_1 < M_2 < \ldots < M_c\}$. Furthermore, let us use $\mathsf{Q}(\mathsf{snr},M) = \mathsf{ber}$ to denote the deterministic mapping between \gls{snr}, $M$ and the expected value of \gls{ber} denoted by $\mathsf{ber}$. The inverse of this mapping for given $M$ is  available and provides  $\mathsf{Q}^{-1}(q_0,M) = \mathsf{snr}_{M,\text{th}}$, where $\mathsf{snr}_{M,\text{th}}$ is the threshold of \gls{snr} for $q_o$ \gls{ber} cross-over for modulation order $M$. Similarly, we have $\mathsf{\gamma}_{M,\text{th}} = \frac{\sigma_n \bar{W} } { \bar{P}} \mathsf{snr}_{M,\text{th}}$.

Both $\mathsf{Q}(.)$ and $\mathsf{Q}^{-1}(.)$ can be empirically derived through sufficient Monte Carlo sampling of the underlying channel distribution.  The network chooses the $M$  that yields the highest spectral efficiency, i.e.,  $\log_2 M$, while stabilizing a target \gls{ber} around $q_o$ for every block, by
\begin{align} \label{eq:modorder}
M = & \arg \max_{m \in \{M_1, \ldots,  M_c\}}  m
\\ \nonumber
&  s.t.  \quad \mathsf{F}_{\gamma | \tilde{\gamma}}\left( \mathsf{\gamma}_{M,\text{th}} \right) \leq \epsilon.
\end{align}
The constraint in \eqref{eq:modorder} aims to curb the chance of \gls{ber} above $q_o$, thus stabilizes the link \gls{ber}.  Note that \eqref{eq:modorder} can readily be extended to a frequency selective channel with multitude of coherence blocks for each frame of communication.
The combination of $M$ and ${W}_t$ then determine the number of bits  that can be supported by the network during block $t$. The parameter $\epsilon \in (0, 1) $ measures the \emph{instability} of the end-to-end binary link  and can be seen as a semantic communication \gls{qos} metric (see discussion in \secref{sec:blueprint}).

\begin{definition}
    A stabilized vector binary symmetric channel $\text{SVBSC}(q_o,\epsilon)$ is a set of parallel \glspl{bsc}  of $\text{BSC}(q)$ where $q$ is stabilized ``around'' $q_o$, that is, $\mathsf{Pr}[ q > q_o] \leq \epsilon$. Similarly, $\text{SVBSC}(q_o,\epsilon, K, \bar{L})$ is a stabilized vector binary symmetric channel of length $K-L$, where the discrete random variable $L \in \{0,\ldots, \bar{L}\}$.
\end{definition}

The network therefore adopts a rate-control process through  mapping $r$, where $L$ ($L \leq \bar{L} < K$) bits are dropped before transmission.
The rate-control function $r$ thus intelligently  \emph{punctures}  $L$ bits out of the $K$ bits received from the source, and provides a $\text{SVBSC}(q_o,\epsilon)$ of length $K - L$ for the remainder of the bits. 

\begin{remark}
    It's worth stressing that such rate-control mechanism is a new paradigm in communication where a portion of the data bits can be intentionally removed by the network. In the conventional communication systems, the network doesn't discard  information bits intentionally---it may deliver them correctly to the receiver as a packet, or drop them altogether, due to packet transmission failure.
\end{remark}

At the receiver side of the network the punctured bits are padded with null values $\varnothing$ (or, 0.5, which contains null information in binary), where the operation is denoted here by $r^{*}$, effectively notifying the decoder $g$ of puncturing of those bits\footnote{In practice, due to the binary interface between network and application on both ends, instead of null bits an accompanying control signal from the network to the application notifies the latter about $L$.}. The maximum number of punctured bits $\bar{L}$ is guaranteed by the network and can be seen as another semantic communication \gls{qos} metric, e.g., corresponding to the highest distortion tolerable by the application. Additionally, a cap on the long-term average of puncturing length over frames can be agreed between the application and the network as another \gls{qos} metric, to guarantee network's best effort in minimizing puncturing (see discussion in \secref{sec:blueprint}).

\begin{remark}
    The uncoded QAM version of $\text{SVBSC}(q_o,\epsilon, K, \bar{L})$, presented above, is only appropriate for the simple case where a channel coherence block and the communication frame align. More commonly, a communication frame may tolerate a latency that spans over multiple coherence blocks in time. Then, use of retransmission techniques, such as hybrid automatic repeat request (HARQ) together with error correction coding can complement the modulation order selection proposed above and improve throughput. Nevertheless, the definition of $\text{SVBSC}(q_o,\epsilon, K, \bar{L})$ remains the same.  For simplicity, here we focused on the uncoded modulation case.
\end{remark}

\begin{figure*}[ht]
\begin{center}
\psfrag{ss}[c][c][1]{$\mathbf{s}$}
\psfrag{ll0}[c][c][1]{$\hat{\mathbf{s}}$}
\psfrag{ll1}[c][c][1]{\scalebox{\textsizescaleM}{$f_{\theta^*_1}$}}
\psfrag{ll2}[c][c][1]{\scalebox{\textsizescaleM}{$f_{\theta^*_2}$}}
\psfrag{ll3}[c][c][1]{\scalebox{\textsizescaleM}{$f_{\theta_i}$}}
\psfrag{ll4}[c][c][1]{\scalebox{.35}{concatenate}}
\psfrag{ll5}[c][c][1]{\scalebox{.35}{concatenate}}
\psfrag{g1}[c][c][1]{\scalebox{\textsizescaleM}{$g_{\phi^*_1}$}}
\psfrag{g2}[c][c][1]{\scalebox{\textsizescaleM}{$g_{\phi^*_2}$}}
\psfrag{g3}[c][c][1]{\scalebox{\textsizescaleM}{$g_{\phi_i}$}}
\psfrag{svbsc1}[c][c][1]{\scalebox{\textsizescaleM}{$\text{SVBSC}(q_o,\epsilon, C_1, 0)$}}
\psfrag{svbsc2}[c][c][1]{\scalebox{\textsizescaleM}{$\text{SVBSC}(q_o,\epsilon, C_2, 0)$}}
\psfrag{svbsc3}[c][c][1]{\scalebox{\textsizescaleM}{$\text{SVBSC}(q_o,\epsilon, C_i, 0)$}}
\psfrag{losssssh3}[c][c][1]{\scalebox{\textsizescaleM}{$\mathcal{L}({\mathbf{s}},\hat{\mathbf{s}})$}}
    \psfrag{delta2}[c][c][1]{\scalebox{\textsizescaleM}{$\Delta{\theta_i}$}}
\psfrag{delta1}[c][c][1]{\scalebox{\textsizescaleM}{$\Delta{\phi_i}$}}
\includegraphics[width=.7\textwidth,keepaspectratio]{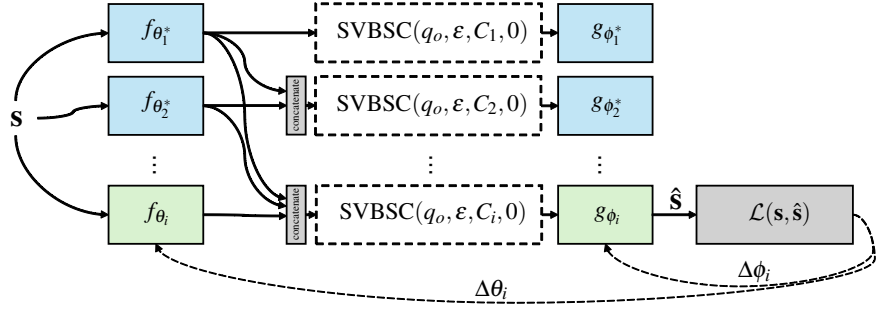}
\caption{Training iteration $i$ for RLACS code, where the trained encoder and decoder modules from previous iterations are frozen (blue color). Dashed arrow lines point to optimization of $f_i$ and $g_i$ based on the loss calculation.}
\label{fig:trainfig}
\end{center}
\end{figure*}

\subsection{Application's Operation: Rateless JSCC}
\label{sec:ratelessjscc}

The  code is designed to minimize the distortion over a binary channel where \gls{ber} is bounded ``around'' $q_o$, and for all the rates in the range $[R_{lo}, R_{hi}]$, where $R_{lo}$ and $R_{hi}$ correspond to $L = \bar{L}$ and $L = 0$ puncturing lengths, respectively.

\begin{definition} \label{def:def1}
A \((K,N, \bar{L}, \mathbf{d})\) binary rateless and lossy joint source-channel code is a tuple of mappings, consisting of an encoder \( f \), a rate control bit-puncturing function $r$, and a decoder \( g \), 
\begin{align}
      f: \mathbb{R}^N\rightarrow \{0,1\}^K, \quad r: \{0,1\}^{K} \rightarrow \{0,1\}^{K-L}  , \quad   g: \{0,1, \varnothing \}^K \rightarrow \mathbb{R}^N 
\label{eq:prb-def}
\end{align}
that  satisfies
\begin{align} \label{Eq:vectorloss}
\forall L \in \{0, \ldots, \bar{L}\} \quad \mathbb{E} \Big( \mathsf{d}({\mathbf{s}}, \hat{\mathbf{s}} ) | L \Big) \leq \mathbf{d}[L],
\end{align}
where the expectation is over the source distribution and channel randomness, 
$\mathsf{d}:  \mathbb{R}^N \times  \mathbb{R}^N \rightarrow [0, \infty)$ is the  distortion function,
$\mathbf{d}$ is a vector of distortion values,
\( \mathbf{u} = f(\mathbf{s}) \), \( \mathbf{x} = r(\mathbf{u}) = \mathbf{u}_{1:K-L} \), \( \mathbf{y} \sim P_{Y|X}(\mathbf{x}) \) is a vector binary channel, 
$\hat{\mathbf{u}} =  \mathbf{y} \Vert \bm{\varnothing}_L$, 
and \( \hat{\mathbf{s}} = g(\hat{\mathbf{u}}) \) (see \figref{fig:system-model}).
\end{definition}

In \defref{def:def1}, $\bm{\varnothing}_L$ denotes a vector of $L$ null values and $\Vert$ denotes concatenation of two vectors. The channel \( \mathbf{y} \sim P_{Y|X}(\mathbf{x}) \) is assumed to be a vector binary channel, as also illustrated in \figref{fig:system-model}. Thus, the channel may in practice include a channel mapping function that maps $\mathbf{x}$ to $M$ complex symbols, passes them through an underlying complex valued channel, and then de-maps the received $M$ complex symbols to binary vector $\mathbf{y}$. Thus, the channel denoted by $P(\mathbf{y}|\mathbf{x})$ comprises the underlying channel and the signal processing functions of the network. 
The SVBSC link from \secref{sec:ratelessradiolink} is an example of such vector binary channel. For ease of notation we omit denoting those constituents of the vector binary channel.

Rateless coding adds another dimension to the performance of source-channel coding which is the \emph{importance} of the bits in reconstruction loss---higher importance is implicitly enforced in bits that are less likely to be punctured, and vice versa, less important information is encoded in bits that are more likely to be punctured. Such unequal importance can be exploited towards semantic and effectiveness encoding. of  As was noted in \secref{sec:priorwork}, other works have already acknowledged this effect in developing \gls{jscc} techniques  under noisy channel conditions \cite{1095155,748707,tung2024multilevelreliabilityinterfacesemantic,bursalioglu2008lossy,bursalioglu2013joint}. Here, we utilize such unequal importance towards rateless-ness of the code, which then enables the network to seamlessly adapt the coding rate based on network state, which differentiates our work.

\begin{definition} \label{def:acheivability}
A rate-distortion mapping vector of $[(\frac{K-l}{N},\mathbf{d}[l])]$ is said to be achievable if there exists a \((K,N, \bar{L}, q_o)\) rateless joint source-channel code as defined in \defref{def:def1}.
\end{definition}

The encoder $f$ operates at the coding rate of $\hat{R} = K/N$, while the mapping $r$ reduces the rate by $(K-L)/K$.
The effective coding rate of $f$ and $r$ mappings is thus $R = (K-L)/M$,   representing the inverse of the compression rate of the lossy and rateless \gls{jscc}. 



The adopted model relaxes the need for a probabilistic model of an adaptive quantizer function for the encoder output, commonly used in variable-rate compression techniques. In fact, the goal of the proposed \emph{rateless} encoding is to commission the network with the rate-control function---a common realization of such function is rate adaptation according to \gls{csi} that is commonly used in wireless communication technologies. 

\begin{remark}
    Effectively, the null bits at the receiver can be seen as bit erasures. From the point of view of decoder $g$, one difference between the flipped bits (ratio $q$ out of $K-L$) and the $L$ null bits is that that the position of the former is unknown while the latter is in a known position (e.g., end of the vector). 
\end{remark}

To the best of author's knowledge, the  rateless JSCC framework and the rate-adaptive and stable link operation presented above, are novel with respect to the literature. In the following section an autoencoder based realization of the rateless JSCC is proposed and its preformance is evaluated over the proposed rate-adaptive and stable link. Investigating the possiblity of using classical compression and channel codes to realize rateless JSCC is for future work.

\section{Rateless and Lossy Autoencoder Channel-Source Codes (RLACS)}
\label{sec:relax}

The design and implementation of a rateless JSCC described in \secref{sec:ratelessjscc} is the subject of this section. We propose an autoencoder based solution with binary latent space for this type of code. The code is designed to operate over the  rate control mechanism  and $\text{SVBSC}(q_o,\epsilon, K, \bar{L})$ channel described in \secref{sec:ratelessradiolink}. We focus on the case of image transmission over a noisy and faded channel, however the proposed solution can be easily extended to semantic communication of other media formats too.

The encoder and decoder mappings  in \eqref{eq:prb-def} are designed as sets of $F$ encoder and decoder mappings $f = \{f_{\theta_1},\ldots, f_{\theta_F} \}$ and $g = \{g_{\phi_1},\ldots, g_{\phi_F} \}$, where $\theta_i$ and $\phi_i$ are  the \gls{dnn} parameter sets that parametrize $f_{\theta_i}$ and $g_{\phi_i}$, respectively. The mappings are designed so that $f_i: \mathbb{R}^N\rightarrow \{0,1\}^{C_i - C_{i-1}}$ and $g_i: \{0,1,\varnothing\}^{C_i} \rightarrow \mathbb{R}^N$, where  $C_0=0$,   $C_1 = K - \bar{L}$ and $C_F = K$, while $F$ and other $C_i$s are design parameters that determine the complexity-performance tradeoff of the training process. Notably, the $i$th decoder takes input that is originated from all the encoders $1,\ldots,i$. Ideally, the $C_i$s are picked to cover all possible puncturing length $L$, i.e., $\mathsf{P}_L(L) \neq 0$.


\subsection{Training}

We first sequentially train an autoencoder ladder structure with $F$ iterations of  training. In the $i$th iteration, the previously trained $i-1$ encoders and decoders parameters  are ``frozen'', that is, they are only used for evaluation and not for training. The $i$th encoder and decoder are trained, while decoder $i$ effectively takes input from encoder $i$ and all the frozen encoders before $i$. Therefore, the training iteration $i$ optimizes the \gls{dnn} parameters as follows. 
\begin{align}\label{eq:opt}
(\theta^*_i, \phi^*_i) = \argmin_{\theta_i, \phi_i} \mathbb{E} \left( \mathcal{L}({\mathbf{s}},\hat{\mathbf{s}}) | \theta^*_{i-1}, \ldots, \theta^*_1 \right),
\end{align}
where $\mathcal{L}({\mathbf{s}},\hat{\mathbf{s}})$ is the  loss measured between the input image $\mathbf{s}$ and its reconstruction $\hat{\mathbf{s}}$, and the expectation is over the training data set and the randomness of the training channel. Note that training iteration $i$ effectively trains the code to minimize the loss  for  puncturing of $L = K - C_i$ in \eqref{Eq:vectorloss}, corresponding to effective coding rate  of $C_i / N$. During the $i$th training iteration, a $\text{SVBSC}(q_o,\epsilon, C_i, 0)$ connects the encoders and decoder $i$. For smooth convergance, during training we assume perfect channel estimate is available to the $\text{SVBSC}$, thus we can set $\epsilon = 0$. The illustration in \figref{fig:trainfig} demonstrates the training phase for the $i$th iteration, where $\Delta{\theta_i}$ and $\Delta{\phi_i}$ represent the flow of loss gradients for parameter optimization.

\subsection{Distortion Minimization}

The loss function utilized for the training process is \gls{mse} reconstruction loss, where
\begin{align}
\mathcal{L}({\mathbf{s}},\hat{\mathbf{s}}) = \frac{1}{N} \sum_{i=1}^{N} \left( s_i - \hat{s}_i \right)^2,
\end{align}
and $N$ is the image size representing the total number of pixels in all  color channels of the image. The inference loss is demonstrated in the following  using the more popular  \gls{psnr}, computed as
\[
\text{PSNR} = 10 \times \log_{10}\left(\frac{\text{1}^2}{\mathcal{L}({\mathbf{s}},\hat{\mathbf{s}})}\right).
\]

\begin{figure}[t]
\begin{center}
\psfrag{xxx0}[c][c][1]{\scalebox{.7}{input CIFAR10 image}}
\psfrag{xxx1}[c][c][1]{\scalebox{.7}{$3 \times 32 \times 32$ pixels}}
\psfrag{ll0}[c][c][1]{\scalebox{\textsizescale}{residual block}}
\psfrag{lll0}[c][c][1]{\scalebox{\textsizescale}{256 output channels}}
\psfrag{ll1}[c][c][1]{\scalebox{\textsizescale}{residual block, $x$-sample}}
\psfrag{lll1}[c][c][1]{\scalebox{\textsizescale}{$y$ output channels}}
\psfrag{k2}[c][c][1]{\scalebox{\textsizescale}{residual block}}
\psfrag{h2}[c][c][1]{\scalebox{\textsizescale}{$y$ output channels}}
\psfrag{ll3}[c][c][1]{\scalebox{\textsizescale}{attention module}}
\psfrag{ll4}[c][c][1]{\scalebox{\textsizescale}{Big Block}}
\psfrag{lll4}[c][c][1]{\scalebox{\textsizescale}{$x=$ down, $y=256$}}
\psfrag{ll5}[c][c][1]{\scalebox{\textsizescale}{residual block}}
\psfrag{lll5}[c][c][1]{\scalebox{\textsizescale}{$\frac{C_i - C_{i-1}}{2}$ output channels}}
\psfrag{ll6}[c][c][1]{\scalebox{\textsizescale}{Big Block}}
\psfrag{lll6}[c][c][1]{\scalebox{\textsizescale}{$x=$ down, $y=256$}}
\psfrag{k1}[c][c][1]{\scalebox{\textsizescale}{Big Block}}
\psfrag{k4}[c][c][1]{\scalebox{\textsizescale}{$x=$ down, $y=\frac{C_i - C_{i-1}}{2}$}}
\psfrag{ll8}[c][c][1]{\scalebox{\textsizescale}{flatten layer}}
\psfrag{ll9}[c][c][1]{\scalebox{\textsizescale}{fully connected}}
\psfrag{lll9}[c][c][1]{\scalebox{\textsizescale}{vector output size $C_i - C_{i-1}$}}
\psfrag{llll}[c][c][1]{\scalebox{\textsizescale}{Sigmoid}}
\psfrag{xxx2}[c][c][1]{\scalebox{.7}{$C_i - C_{i-1}$ coded bits}}
\psfrag{k3}[c][c][1]{\scalebox{.7}{Big Block $x$, $y$}}
\psfrag{yyy0}[c][c][1]{\scalebox{.7}{(channeled) coded bits}}
\psfrag{yyy1}[c][c][1]{\scalebox{.7}{vector size $C_i$}}
\psfrag{g0}[c][c][1]{\scalebox{\textsizescale}{fully connected, 3D output}}
\psfrag{uu0}[c][c][1]{\scalebox{\textsizescale}{ size of $( C_i \times z) \times 8 \times 8$}}
\psfrag{g1}[c][c][1]{\scalebox{\textsizescale}{spatial reshape}}
\psfrag{uu1}[c][c][1]{\scalebox{\textsizescale}{leaky ReLU}}
\psfrag{g3}[c][c][1]{\scalebox{\textsizescale}{Big Block}}
\psfrag{uu3}[c][c][1]{\scalebox{\textsizescale}{ $y= C_i \times z$}}
\psfrag{g4}[c][c][1]{\scalebox{\textsizescale}{Big Block}}
\psfrag{uu4}[c][c][1]{\scalebox{\textsizescale}{$y= 2 \times C_i \times z$}}
\psfrag{h0}[c][c][1]{\scalebox{\textsizescale}{Big Block}}
\psfrag{ss}[c][c][1]{\scalebox{\textsizescale}{$x = $ up, $y= 2 \times C_i \times z$}}
\psfrag{g6}[c][c][1]{\scalebox{\textsizescale}{Big Block}}
\psfrag{uu6}[c][c][1]{\scalebox{\textsizescale}{$x = $ up, $y=  C_i \times z$}}
\psfrag{g7}[c][c][1]{\scalebox{\textsizescale}{attention module}}
\psfrag{g8}[c][c][1]{\scalebox{\textsizescale}{Big Block}}
\psfrag{uu8}[c][c][1]{\scalebox{\textsizescale}{$y= \frac{1}{2} \times C_i \times z$}}
\psfrag{n4}[c][c][1]{\scalebox{\textsizescale}{residual block}}
\psfrag{m4}[c][c][1]{\scalebox{\textsizescale}{$3$ output channels}}
\psfrag{nn}[c][c][1]{\scalebox{\textsizescale}{Sigmoid}}
\psfrag{yyy2}[c][c][1]{\scalebox{.7}{reconstructed image}}
\psfrag{yyy3}[c][c][1]{\scalebox{.7}{$3 \times 32 \times 32$ pixels}}
\includegraphics[width=\archfigscale\columnwidth,keepaspectratio]{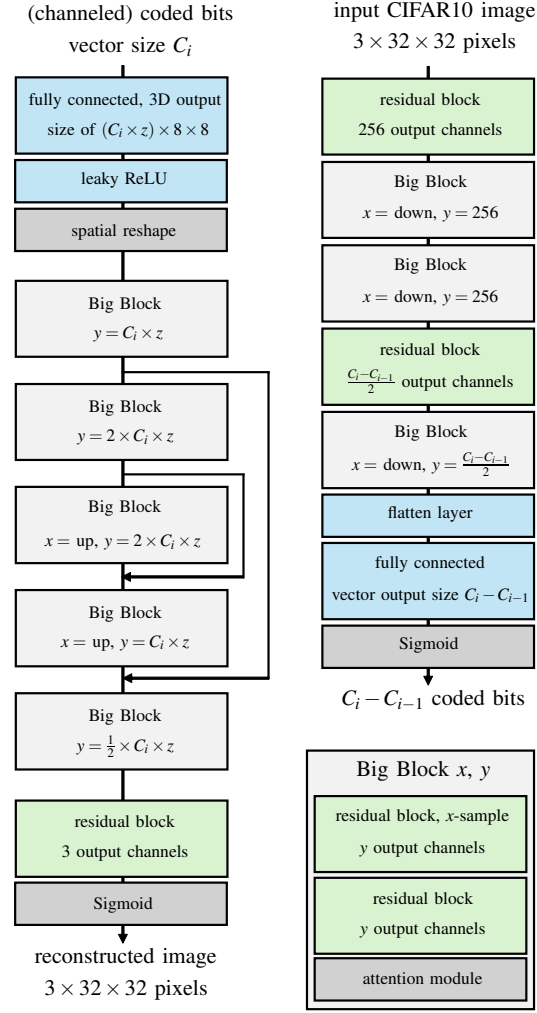} 
\caption{Architecture of the \gls{dnn} used for $f_{\theta_i}$  and  $g_{\phi_i}$ are shown on the right and the left sides, respectively. The dimensions of the encoder and decoder layers adapts to $C_i - C_{i-1}$ and $C_i$, to ensure a sufficient capacity in the network architecture. Additionally, in the decoder structure,  layer dimensions first expand through the middle of the network and then contract, using a factor $z = \min (1, 256/C_i)$. To simplify the illustration, a ``Big Block'' is introduced in the bottom right corner, which takes a variable $x$ to choose between up-sampling and down-sampling (stride 2), if either is used. In the decoder, skip connections retain intermediate outputs from earlier layers and integrate them into later layers. The retained outputs are resized to match the current layer's resolution and combined with its features. An attention module then selects the most important information from the combined features.
}
\label{fig:encdec}
\end{center}
\end{figure}

\begin{figure*}[ht]%
    \centering
    %
    \begin{minipage}[t]{0.45\linewidth}
        \begin{algorithm}[H]
        \caption{RLACS encoding}
        \label{alg:enc}
        \begin{algorithmic}[1]
        \REQUIRE Sample image $\mathbf{s}$, $\Theta^* = \{\theta^*_1, \ldots, \theta^*_F\}$
        \STATE $\mathbf{u} = [ \ ]$
        \FOR{$i = 1$ \TO $F$}
            \STATE $\mathbf{u}=\text{queue\_push}(\mathbf{u},f_{\theta^*_i}(\mathbf{s}))$
        \ENDFOR
        \RETURN $\mathbf{u}$
        \end{algorithmic}
        \end{algorithm}
    \end{minipage}
    \hfill
    %
    \begin{minipage}[t]{0.51\linewidth}
        \begin{algorithm}[H]
        \caption{RLACS decoding}
        \label{alg:dec}
        \begin{algorithmic}[1]
        \REQUIRE $K$, $L$, $\hat{\mathbf{u}}$, $\Phi^* = \{\phi^*_1, \ldots, \phi^*_F\}$,  $\{C_1, \ldots, C_F\}$
        \STATE $\mathbf{r}=\hat{\mathbf{u}}_{1:K-L}$  
        \STATE $i \gets \argmin_{j} \{ C_j \mid C_j \geq K-L \} $
        \IF{$C_i > K-L $}
            \STATE $\mathbf{r}=\text{queue\_push}(\mathbf{r},\mathbf{0.5}_{C_i - K + L})$
        \ENDIF
        \RETURN $\hat{\mathbf{s}} = g_{\phi^*_i}({\mathbf{r}})$
        \end{algorithmic}
        \end{algorithm}
    \end{minipage}
    %
\end{figure*}

\subsection{Evaluation and Inference}
During evaluation, the encoder output is quantized to binary by simple thresholding of the Sigmoid layer output around $0.5$.
The trained \gls{relax} code is tested over $\text{SVBSC}(q_o,\epsilon, K, \bar{L})$. The application source  encodes using $f = \{f_1,\ldots, f_F \}$  at maximum rate of $K/N$ (see \Algref{alg:enc}), while the application destination, being aware of the puncturing length, picks the corresponding decoder from the set $\{g_1,\ldots, g_F \}$ and performs decoding over the received bits from the network (see \Algref{alg:dec}).

\begin{remark}\label{rem:vectorepsilon}
    In practice, the proposed solution can be designed for a set of $q_i$ and $\epsilon_i$ values related to $i \in \{1,\ldots,F\}$. Such setting will introduce more degrees of flexibility for optimizing the joint venture between the network and the application. In the following we foucs on $q_i = q_o$ and $\epsilon_i = \epsilon$ for all $i$.
\end{remark}


\subsection{An Example RLACS Code}
\label{sec:examplerlacs}

In the following we present one example of a \gls{relax} code. We use \glspl{dnn}, more specifically  residual attention network architecture \cite{wang2017residual} with binary latent space to embody the encoders and decoders. To convert the real-valued  output of the  encoder \glspl{dnn} into a binary latent space during trainig,  we use the  variational inference for Monte Carlo objectives (VIMCO) estimator  proposed in \cite{mnih16_variat_monte_carlo}   to estimate the gradient of the loss function with respect to the \glspl{dnn} weights. Notably, 50 Monte Carlo samples are used for each training step. The architectures  used for $f_{\theta_i}$  and  $g_{\phi_i}$ are illustrated in \figref{fig:encdec}.

The models  are trained over  $\text{SVBSC}(q_o,\epsilon, C_i, 0)$ training channel with $q_o = 0.05$ and $\epsilon = 0$. The choice of $q_o$ can arguably affect the performance. For example, a lower $q_o$ could improve  performance of the autoencoder module and increase convergence rate, while in turn it reduces the spectral efficiency of the channel. Therefore, there exists  a clear empirical tradeoff for $q_o$ optimization---in this work, we do not exercise that optimization and refer it to future research.

All models are trained with the similar settings: a base learning rate of $10^{-3}$ to $10^{-5}$, an  Adam optimizer with $\beta_1 = 0.9$, $\beta_2 = 0.999$, decay $= 0$, a batch size of 320  and training epochs from 500 to 2000 (depends on model size). A learning rate scheduler reduces learning rate by factor of $0.5$ with 10 epochs patience, and stops the training after 50 epochs patience, where the patience for both is based on invariability of validation loss that is based on  a  subset of the data disjoint from the training and testing set. The GradScaler is used to facilitate mixed-precision training \cite{paszke2019pytorch} by dynamically scaling gradients with an initial scale of  $2^{16}$, adjusting it using a growth factor of $2$ and a backoff factor of $0.5$ every $2000$ steps.

We utilize the CIFAR10 image dataset \cite{krizhevsky09_learn_multip_layer_featur_tiny_images} where, out of the 60K images, 40K and 10K disjoint subsets are  selected for training and evaluation, respectively. Each image has three color channels amounting to total of $3 \times 32 \times 32$ real valued pixels, resulting in $N = 3072$. The network is assumed to utilize $W = 128$ channel uses, resulting in an effective pixel real-value to channel real-value compression ratio of $1/24$. 

\subsection{Evaluation Results}

Two testing channels are explored to test the example \gls{relax} code proposed in \secref{sec:examplerlacs}. In both cases, we assume a block fading channel model that follows Rician distribution  with line-of-sight ratio of $20$ dB.

\paragraph*{Perfect CSI}  The test channel is block fading with perfect \gls{csi} at the transmitter and the receiver ends of the network. In effect, this resembles an \gls{awgn} channel with varying but known SNR.

\paragraph*{Imperfect CSI} To study the effect of imperfect CSI (and $\epsilon$) on the performance, we assume a  block fading channel with imperfect  \gls{csi}, where the Gaussian channel estimation noise variance $\sigma_e$ is proportional to \gls{snr}, according to \cite{1193803,9206086}, as  $\sigma_e = \frac{1}{1+n_p \cdot \mathsf{snr}}$, with $n_p$ denoting the number of pilot symbols used per coherence block for channel estimation and $\mathsf{snr}$ is the wireless channel \gls{snr}. We set $n_p = 10$.

\begin{figure}[t!]
\begin{center}
\psfrag{ber}[c][c][1]{\scalebox{.8}{average bit error rate}}
\psfrag{snr}[c][c][1]{\scalebox{.8}{channel signal to noise ratio [dB]}}
\psfrag{Xxxxxxxx1Xxxxxxxx1}[l][l][1]{\scalebox{\textsizescale}{$\mathsf{Q}(\mathsf{snr},M)$, uncoded QAM}}
\psfrag{data1}[l][l][1]{\scalebox{\textsizescale}{$q_o = 0.05$}}
\psfrag{Xxxxxxxx3Xxxxxxxx1}[l][l][1]{\scalebox{\textsizescale}{perfect \gls{csi}}}
\psfrag{X0}[l][l][1]{\scalebox{\textsizescale}{imperfect CSI, $\epsilon = 0.01$}}
\psfrag{X1}[l][l][1]{\scalebox{\textsizescale}{imperfect CSI, $\epsilon = 0.05$}}
\psfrag{X2}[l][l][1]{\scalebox{\textsizescale}{imperfect CSI, $\epsilon = 0.1$}}
\includegraphics[width=\imagescalesize\columnwidth,keepaspectratio]{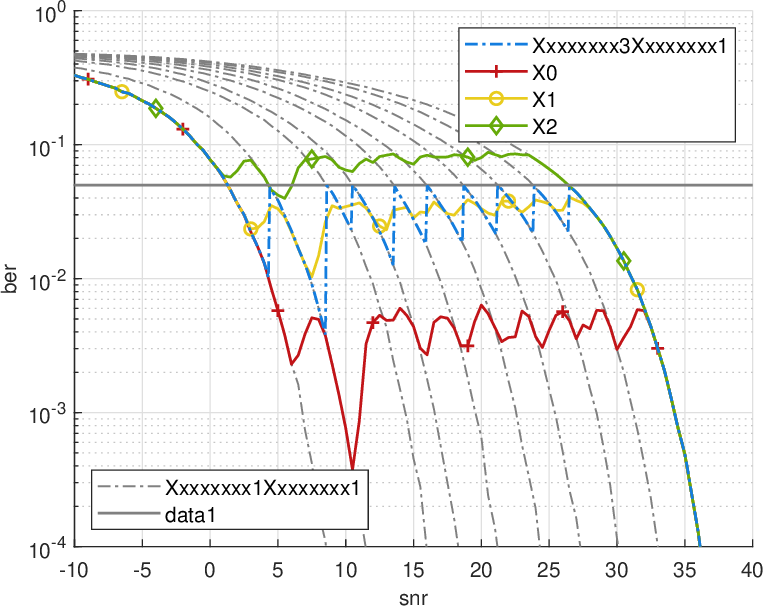}
\caption{Average bit error rate experienced over the  testing channels $\text{SVBSC}(q_o=0.05,\epsilon, 1280, 1152)$ (top legend) against the underlying wireless channel \gls{snr} with $\bar{W}=128$. The uncoded QAM curves cover $M = {2, 4, 8, 16, 32, 64, 128, 256, 512, 1024}$ modulation orders (ordered from left to right). In the case of perfect \gls{csi}, \gls{ber} can be perfectly kept below $q_o$. In case of imperfect \gls{csi}, $\epsilon$ determines the average \gls{ber}. A tight $\epsilon$ of $0.01$ for instance, results in average \gls{ber} almost one order of magnitude below $q_o$. Average \gls{ber} is not in itself  an important measure---more critical metric is the average reconstruction loss over the randomness of \gls{ber}, when \gls{csi} is imperfect.}
\label{fig:bervssnr}
\end{center}
\end{figure}

For these two test channels, \figref{fig:bervssnr} shows the average  \gls{ber} against channel \gls{snr}. The case of perfect CSI follows the \gls{ber} of the uncoded QAM modulation. For the imperfect CSI case, the average \gls{ber} naturally depends on the configured $\epsilon$. Notably, a tighter constraint on $\epsilon$ results on  a lower average \gls{ber}. However, as shown in \figref{fig:etavssnr}, this in turn results in lowering the espectral efficiency of the link (i.e., increased chance of using a lower modulation order). This demonstrates the tradeoff between stability and efficiency in a $\text{SVBSC}$ link. We test our RLACS code over this tradeoff to see the effect of instability parameter $\epsilon$ on image reconstruction quality too.

Three main JSCC codes are tested over the test channels, where all three are based on the  \gls{dnn} architectures in \figref{fig:encdec}. 
\begin{itemize}
    \item \textbf{Code 1},  $F = 1$ and $C_1 = 1280$: corresponds to a benchmark where the code is trained for maximum spectral efficiency of 10 bpcu, but the code is not rateless, i.e., it is not specifically optimized with puncturing effect from the network in mind. 
    \item \textbf{Code 2}, $F = 2$, and $C_i-C_{i-1} = 640$ for all $i$: is the proposed \gls{relax} code, thus, is rateless, and is optimized  for code lengths of $i \times 640$ for $i \in \{1, 2\}$. 
    \item \textbf{Code 3}, $F = 10$, and $C_i-C_{i-1} = 128$ for all $i$: is also the proposed \gls{relax} code, but with $C_i$s adjusted such that the puncturing lengths (resulted from choice of modulation order) aligns perfectly as $C_i \leftrightarrow \bar{W} \cdot \log_2(M_i)$, thus providing finer granularity of rateless-ness compared to the above two codes. 
\end{itemize}
Notably, the codes are sized to the maximum spectral efficiency of the test channel with $M_c = 1024$ QAM, which is $\bar{W} \cdot \log_2(M_c) = 1280$ bits---that is, $C_F = 1280$ for all three codes. It's worth noting that a RLACS code  can be easily extended to cover higher modulation orders (and higher spectral efficiency) by adding more encoder and decoder pairs to the ensemble, without the need to retrain the previous encoder and decoder pairs.

Code 1 is optimized for the maximum rate (10 bpcu) with no constraints related to lower rates and thus its performance should be seen as a the upper bound of the performance for the other two codes at 10 bpcu rate, i.e., in high \gls{snr}. Nevertheless, the experimental results in \figref{fig:psnrcodes}  show that the gap between the three curves in the high rate/SNR region is in fact negligible, thus empirically demonstrating  optimality of the example \gls{relax} code with large $F$.

\subsection{Discussion of the Results and Take-Away Points}
\label{sec:discussionofresults}

Let us now delve into the details of the presented results and discuss the important points.

\paragraph*{Ralteless coding significantly improves performance, especially in low SNR} As \gls{snr} decreases, and the modulation order that satisfies \eqref{eq:modorder} decreases, more of the bits has to be punctured off of the  packet. Code 1 is optimized for 10 bpcu rate and is not rateless.  Thus, while this code  perform optimally well at high \gls{snr} it is sub-optimal for low \gls{snr}. The gap between the performance of code 1 and code 3 illustrates the effect of rateless JSCC coding in maintaining efficient performance against all SNR values, without the need to exchange CSI knowledge with the application. Notably,  code 3 outperforms code 1 in \figref{fig:psnrcodes} by more than $5$ dB in reconstruction quality, at low channel SNR, emphasizing the importance of rateless JSCC design. The rateless characteristics of a code in effect enforces higher importance in bits that are less likely to be punctured, and vice versa. 

\begin{figure}[t!]
\begin{center}
\psfrag{eta}[c][c][1]{\scalebox{.8}{average spectral efficiency [bpcu]}}
\psfrag{channel SNR [dB]}[c][c][1]{\scalebox{.8}{channel signal to noise ratio [dB]}}
\psfrag{Xxxxxxxx1Xxxxxxxx1}[l][l][1]{\scalebox{\textsizescale}{perfect CSI}}
\psfrag{X0}[l][l][1]{\scalebox{\textsizescale}{imperfect CSI, $\epsilon = 0.01$}}
\psfrag{X1}[l][l][1]{\scalebox{\textsizescale}{imperfect CSI, $\epsilon = 0.05$}}
\psfrag{X2}[l][l][1]{\scalebox{\textsizescale}{imperfect CSI, $\epsilon = 0.1$} }
\includegraphics[width=\imagescalesize\columnwidth,keepaspectratio]{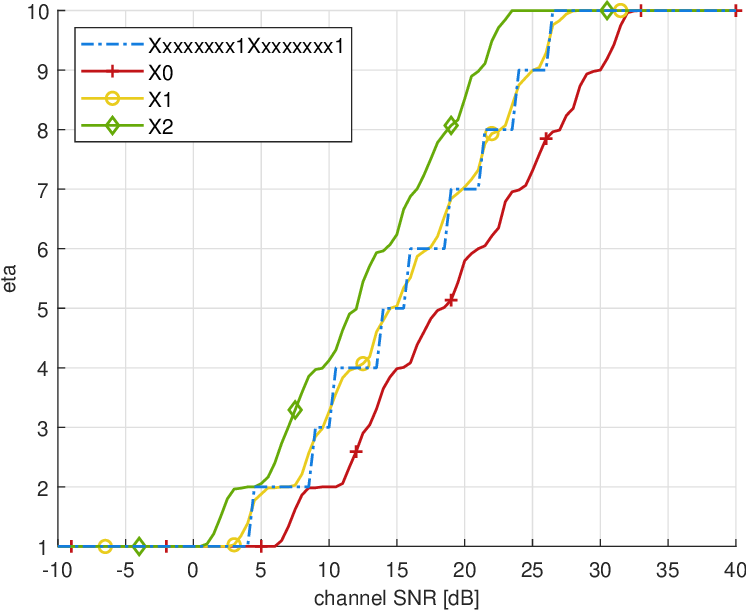}
\caption{Average spectral efficincy of the curves in \figref{fig:bervssnr}. With $\epsilon = 0.05$ the average spectral efficiency, and the average \gls{ber} closely follow the case of perfect \gls{csi}.}
\label{fig:etavssnr}
\end{center}
\end{figure}

\paragraph*{Rateless coding performs  in  harmony with the rate-adaptive and stable link for semantic communication} The impact of stability parameter $\epsilon$ on the rateless coding performance is illustrate in \figref{fig:psnr}, for code 3. The closest performance to the case of perfect CSI, with fraction of a dB gap, is achieved  with $\epsilon = 0.05$. Incidentally, $\epsilon = 0.05$ also provides the closest spectral efficiency to the case of  perfect CSI in \figref{fig:etavssnr}. Note that we have only explored a few choices of $\epsilon$ values here (mainly selected based on the average BER performance those offer in \figref{fig:bervssnr}), and a more careful optimization of $\epsilon$ may further improve the performance of our proposed solutions.

\paragraph*{Granularity of RLACS code design is critical in low SNR} Granularity of RLACS code is controlled by $C_i$ parameters. Ideally, those design parameters must match the possible puncturing lengths, so that at each puncturing length distortion minimization is guaranteed---code 3 is designed so. When the granularity is lowered, e.g., in case of code 2, the performance is affected, especially in low SNR. Notably, code 2 and code 3 have comparable performance around the point of 5 bpcu, i.e., around 14 dB channel SNR. That is the point where code 2 experiences zero puncturing thanks to its first decoder. As channel SNR grows beyond 14 dB, code 2 again falls below code 3 in performance until the two curves meet again at 10 bpcu region, i.e., right side of \figref{fig:psnrcodes}. This further stresses the importance of granularity in RLACS code design. In practical system design, the possible range of puncturing length can be pre-agreed on between the network and the application, e.g., through standard specified look-up tables. The application can then make sure to match those values while optimizing the code.

\begin{figure}[t!]
\begin{center}
\psfrag{ber}[c][c][1]{\scalebox{.8}{image reconstruction PSNR [dB]}}
\psfrag{snr}[c][c][1]{\scalebox{.8}{channel signal to noise ratio [dB]}}
\psfrag{Xxxxxxxx2Xxxxxxxx1Xxxxxxxx0}[l][l][1]{\scalebox{\textsizescale}{Code 1, not rateless}}
\psfrag{X1}[l][l][1]{\scalebox{\textsizescale}{{Code 2}, rateless with low granularity}}
\psfrag{X2}[l][l][1]{\scalebox{\textsizescale}{{Code 3}, rateless with ideal granularity}}
\psfrag{X3}[l][l][1]{\scalebox{\textsizescale}{imperfect CSI, $\epsilon = 0.1$} }
\includegraphics[width=\imagescalesize\columnwidth,keepaspectratio]{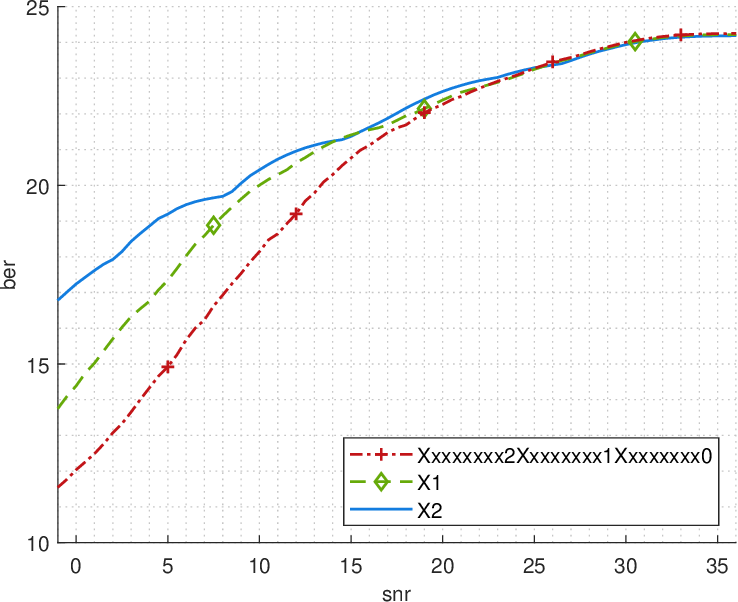}
\caption{Image reconstruction  in PSNR for the three  codes, tested with perfect CSI testing channel.}
\label{fig:psnrcodes}
\end{center}
\end{figure}

\paragraph*{RLACS performs strongly against channel SNR variation}   \gls{relax} code can be  compared with the solutions in \cite{bourtsoulatze19_deep_joint_sourc_chann_codin} and \cite{tung2024multilevelreliabilityinterfacesemantic}. We refer to  \cite[Figure~10]{tung2024multilevelreliabilityinterfacesemantic}, where those two solutions are tested for a similar compression ratio of $1/24$ over \gls{awgn} channel (comparable to our testing channel with perfect CSI). Most notably, code 3 performs on par with Split-JSCC of \cite{tung2024multilevelreliabilityinterfacesemantic}, while in contrast, RLACS relaxes the need for code optimization over several training SNRs, and, only requires JSCC operation in the application, significantly reducing the optimization complexity. There is however a gap in performance of RLACS compared to DeepJSCC \cite{bourtsoulatze19_deep_joint_sourc_chann_codin} which is rooted in the assumption of a binary interface between the application and the network in our proposal. Performance of RLACS, thanks to the rateless operation, gracefully degrades as \gls{snr} decreases, and avoids the \emph{cliff effect} and \emph{leveling-off effect} associated to separate source and channel coding  reported in \cite{gunduz2024joint}, while maintaining separation of design and optimization of the network and application, intact. In fact, looking at code 1 performance in \figref{fig:psnrcodes}, one can postulate that those effects are mainly the result of packet-ized and error-free approach to communication of bits in conventional separate source-channel coding (see further discussion in \secref{sec:blueprint}).

\begin{figure}[t!]
\begin{center}
\psfrag{ber}[c][c][1]{\scalebox{.8}{image reconstruction PSNR [dB]}}
\psfrag{snr}[c][c][1]{\scalebox{.8}{channel signal to noise ratio [dB]}}
\psfrag{Xxxxxxxx1Xxxxxxxx0}[l][l][1]{\scalebox{\textsizescale}{perfect CSI}}
\psfrag{X1}[l][l][1]{\scalebox{\textsizescale}{imperfect CSI, $\epsilon = 0.01$}}
\psfrag{X2}[l][l][1]{\scalebox{\textsizescale}{imperfect CSI, $\epsilon = 0.05$}}
\psfrag{X3}[l][l][1]{\scalebox{\textsizescale}{imperfect CSI, $\epsilon = 0.1$} }
\includegraphics[width=\imagescalesize\columnwidth,keepaspectratio]{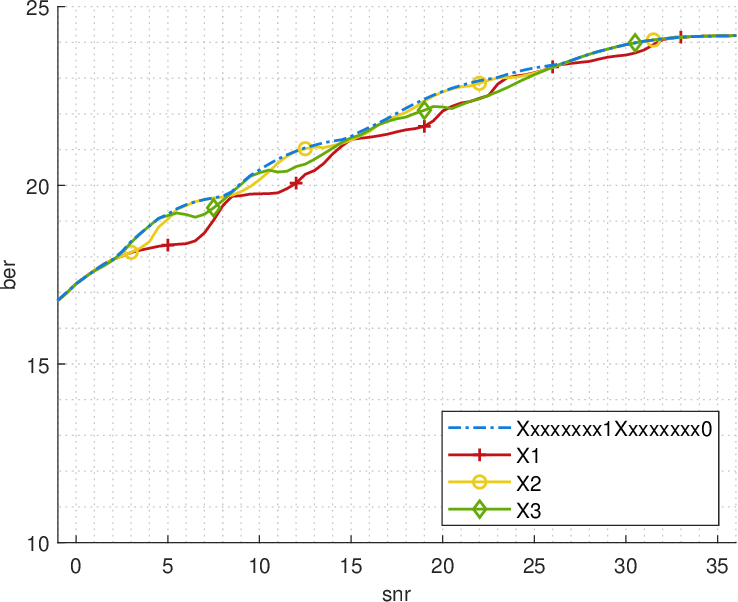}
\caption{Image reconstruction of the testing process in PSNR for code 3. The case of perfect CSI (also present in \figref{fig:psnrcodes}) provides a reference point to demonstrate the performance loss due to channel estimation error. Notably, the gap caused by imperfect CSI is a fraction of a dB when $\epsilon = 0.05$.}
\label{fig:psnr}
\end{center}
\end{figure}

\paragraph*{Modulation order granularity is critical for smooth rate-adaptation in SVBSC link} As shown in \figref{fig:bervssnr}, the available QAM modulation orders in \gls{5g} system (i.e., 2--1024) provide a limited range of \gls{snr} for a given target BER $q_o$. At SNR values lower and higher than that range, the experience BER will be higher and lower than the target, respectively. Extending that range is easily possible by adding higher modulation orders to cover beyond $ M = 1024$, and by adding channel codes with $<1$ code rate to $M=2$ modulation to cover below $M=2$. The RLACS code granularity (i.e., $C_i$ values) can then  be accordingly adjusted. Alternatively, as noted earlier in \remref{rem:vectorepsilon}, different  $q$ and $\epsilon$ values can be used for the first and the last encoder and decoder pair, compared to the rest, to match the actual BER in \figref{fig:bervssnr}.

\paragraph*{JSCC can be fully performed by the application, without the knowledge of the network's state} Exchanging network's state with the application can become a hurdle in practice, preventing the code to perform at its optimal rate. With rateless JSCC this issue is efficiently resolved by commissioning the network with the rate-control responsibility according to network's state, while ensuring optimal code design for potential puncturing of the codeword bits by the network. Unlike the works in \cite{bourtsoulatze19_deep_joint_sourc_chann_codin} and \cite{tung2024multilevelreliabilityinterfacesemantic}, where training (optimizing) a code for different SNR values is exercised to match the training and testing channels, our proposed rateless JSCC approach relaxes the need for the application to worry about the \emph{channel}.

\begin{figure}[t]
\begin{center}
\includegraphics[width=\columnwidth,keepaspectratio]{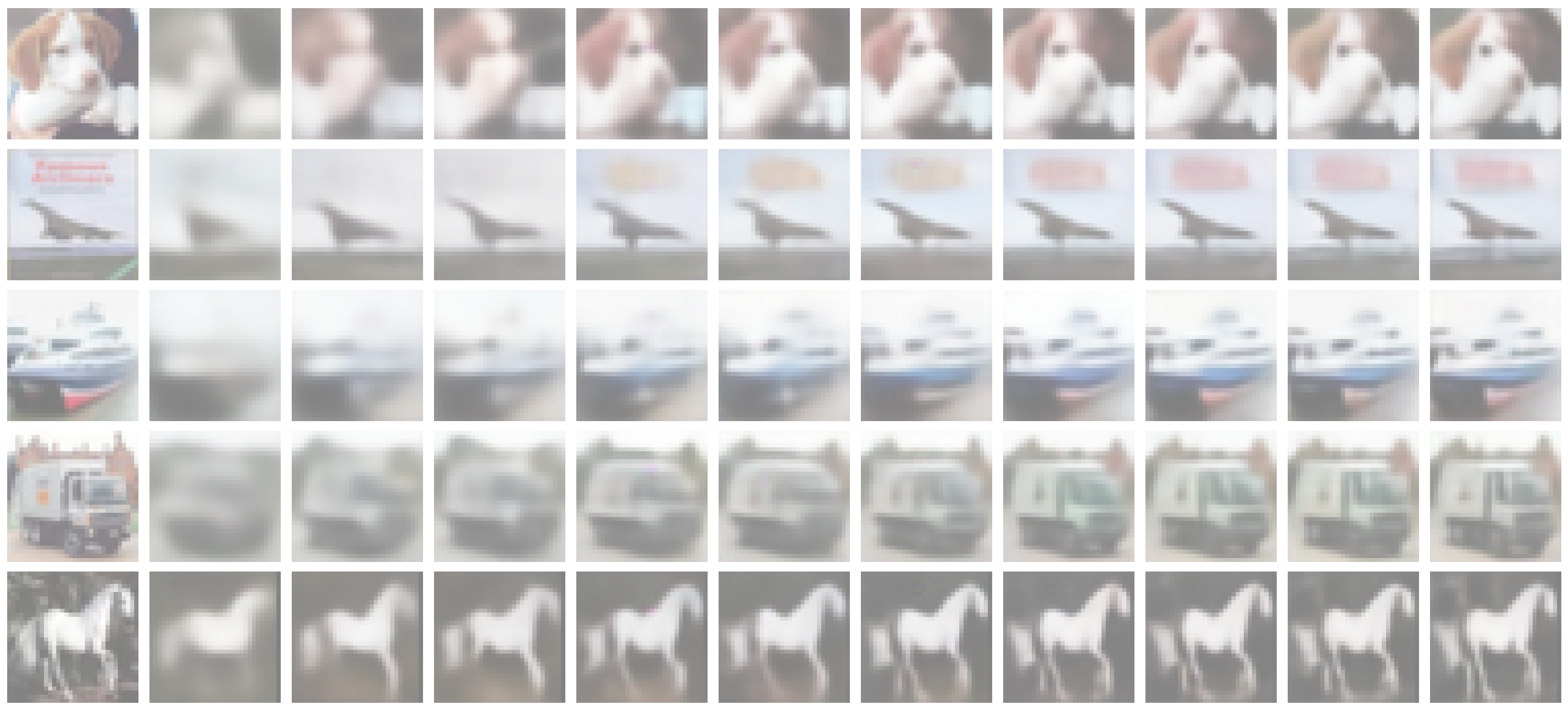}
\setlength{\unitlength}{\columnwidth} 
\begin{picture}(1,0)  
    \put(0.045,0){\makebox(0,0){\scalebox{\LabelFontSize}{Original}}}
    \put(0.135,0){\makebox(0,0){\scalebox{\LabelFontSize}{$M = 2$}}}
    \put(0.225,0){\makebox(0,0){\scalebox{\LabelFontSize}{$M = 4$}}}
    \put(0.315,0){\makebox(0,0){\scalebox{\LabelFontSize}{$M = 8$}}}
    \put(0.405,0){\makebox(0,0){\scalebox{\LabelFontSize}{$M = 16$}}}
    \put(0.495,0){\makebox(0,0){\scalebox{\LabelFontSize}{$M = 32$}}}
    \put(0.585,0){\makebox(0,0){\scalebox{\LabelFontSize}{$M = 64$}}}
    \put(0.675,0){\makebox(0,0){\scalebox{\LabelFontSize}{$M = 128$}}}
    \put(0.765,0){\makebox(0,0){\scalebox{\LabelFontSize}{$M = 256$}}}
    \put(0.855,0){\makebox(0,0){\scalebox{\LabelFontSize}{$M = 512$}}}
    \put(0.945,0){\makebox(0,0){\scalebox{\LabelFontSize}{$M = 1024$}}}
\end{picture}
\caption{Progressive coding effect of the proposed rateless JSCC is evident in the examples illustrated here. Original image is compared against the reconstructions using Code 3, where the modulation order used for the communication channel is noted below each column. As modulateion order increases more redundancy becomes available to the decoder, i.e., puncturing length $L$ becomes smaller, hence, the reconstruction quality progressively improves.}
\label{fig:comparison}
\end{center}
\end{figure}

\paragraph*{Progressive performance improvement as rate increases} As shown in \figref{fig:comparison}, the proposed rateless JSCC framework, and the DNN-based realization of it in RLACS, can be seen as \emph{progressive} JSCC coding solutions. Conventional progressive coding is used for compression purposes without dealing with errors, or in combination  with channel coding to recover from errors too \cite{sherwood1997progressive}. Rateless JSCC is thus, to our knowledge, the first true progressive JSCC solution, and is able to  deal with channel errors while being  unaware of the channel state.

\paragraph*{A systematic and practical framework for code design and semantic communication}
Importantly, \gls{relax} code is built on a systematic approach with practical considerations taken into account. \gls{relax} code doesn't require to be designed separately for each \gls{snr} operation point and can be optimized agnostic to the undelying channel. This is a major advantage of general separate source-channel coding solutions which was missing in joint source-channel codings. With rateless JSCC and rate-adaptive stable link operation, the performance benefit of JSCC coding is offered while maintaining the modular design and optimization benefits of conventional communication networks, allowing the two to evolve and be optimized independently.

\paragraph*{Additional points to consider towards future work}
\begin{itemize}
    \item While autoencoders show good potential for error correction coding, optimization  and training convergance for them proves complex, especially, when the error rates are too high. That suggests targetting a low $q_o$ in designing \gls{relax} codes. A lower BER requires a  higher \gls{snr}, thus,  there is a clear tradeoff to be explored in designing \gls{dnn}-based rateless JSCCs. For example, it may be better to do additional error correction using well-optimized channel codes such as \gls{ldpc} or Polar codes, either at the network or at the application.  The role of such channel code is not to correct all the errors, but to reduce and stabilize the \gls{ber}. This way, the power of classic channel codes in error correction will be combined with the versatility of \glspl{dnn} in semantic \gls{jscc}. As the reader may be picturing it at this point, through these ideas we are exploring the \emph{spectrum} of design options between \emph{fully-separate} and \emph{fully-joint} source-channel coding. 
    \item In practical systems, the network and the application have limited exchange of information. While the interface between the two can improve, and the information exchanged between them can become richer (e.g., see discussion in \secref{sec:blueprint}), it will remain impractical to share real-time information about the channel state with the application. Thus, the joint venture between the network and application must consider this pragmatic separation of responsibilities: network stabilizes the end-to-end bit pipe in one dimension (e.g., \gls{ber}) while remain stochastic in the other dimension (size of the pipe, i.e., a random puncturing rate), and the application tries to maximize its utilization of the stabilized end-to-end channel, knowing that behavior. 
\end{itemize}


\section{Blueprint of a 6G Semantic Communications System }
\label{sec:blueprint}

This section investigates the semantic communication problem form a system design perspective through the lens of practicality. Next generations of communication systems, especially the \gls{6g} networks, are anticipated to see more applications of semantic  type \cite{saad2024foundations}, including spatial video and mixed/augmented reality. In the following, first we provide a sketch of the system design differences between  technical communication and  semantic and effectiveness communication. Then, we project that sketch over the existing communication network systems, the \gls{5g} in particular, and discuss the changes that need to happen. Finally, we provide a list of open research problems and development challenges that require further investigations. To that end, the list will include items related directly to the blueprint  provided in this paper, and  items discussed by other works in the literature.

\subsection{Sketch Summary}
\label{sec:sketch}
The most critically new aspect of the semantic communication system paradigm is the assumption that the semantic applications are \emph{error-tolerant}---that is, the application can take in bit errors and bit erasures in the communicated packet and turn that into tolerable \emph{loss} or \emph{distortion} in the semantic performance indicator. In contrast, the existing networking solutions are mainly designed to handle error-free communication. As a result,  current applications that most resemble  Weaver's definition of semantic and effectiveness communication in \cite{shannon1998mathematical}---e.g., multimedia communication in form of image and video---operate ``over-the-top'' on the existing networking solutions with minimal or zero cooperative interactions with the network. Consequently, from the application's perspective, the end-to-end link is error-free, but may constitute packet  erasures that are visible to the application. A demonstrative example of this is video communication over \gls{5g} systems, where video codecs deploy \emph{error concealment} techniques to deal with partial or complete erasure of video frames \cite{aign1995temporal}.  
What the future network systems can add to this, especially in the context of \gls{6g}, is mainly to enable bit-error-tolerance in the end-to-end system design, where the network allows passage of erroneous bits to the application and the application learns how to deal with those errors. The catch is that the position  of those errors  are not visible to the application (as opposed to the position of packet erasures in existing approach described above), but the rate of error can be realistically anticipated and stabilized. As discussed in the previous sections of this paper, that makes the end-to-end link into a \emph{stabilized bit pipe}, or a stabilized vector binary channel, from the application's perspective.

\begin{figure*}[h]
\begin{center}
\includegraphics[width=.85\textwidth,keepaspectratio]{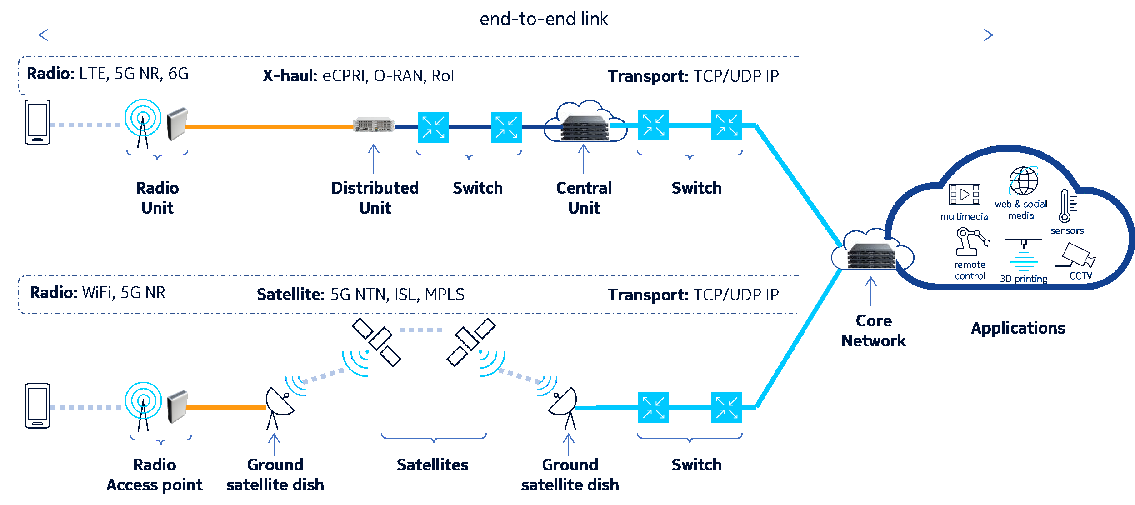}
\caption{Illustration of the typical constituents of an end-to-end communication link, with an example terrestrial link at the top and an example non-terrestrial link at the bottom. The radio, X-haul, satellite, and transport protocols must be able to interoperate with each other and across different vendors.}
\label{fig:e2e}
\end{center}
\end{figure*}

\subsection{Blueprint of Semantic Communications System design}

Existing literature on semantic communication  relies on  \gls{jscc} solutions that fuse the application and the network treatments of a semantic packet. This includes solutions for dismantling   the   binary interface  that separates application's source coding and network's channel coding \cite{bourtsoulatze19_deep_joint_sourc_chann_codin}, or, solutions that  work around that interface \cite{tung2024multilevelreliabilityinterfacesemantic}. For  practical reasons, straightforward \gls{jscc} solutions  won't be feasible, most notably,  due to interoperability requirement among network and application vendors, and interoperability among different functions and hops of an end-to-end network link. For example, accepting a quantized (binary) interface between network and application 
appears necessary\footnote{In very special applications, usually dealing with local traffic,  such  quantization requirement can be relaxed, enabling channel-adaptive \gls{jscc} solutions, such as in\cite{bourtsoulatze19_deep_joint_sourc_chann_codin}.}.

In the most common communications settings,  the end-to-end link includes several constituent hops with variety of attributes: fiber and cable links are typically more reliable with static behavior while wireless links can experience highly dynamic channel state---see the two examples in \figref{fig:e2e}. The end-to-end bit pipe in application's perspective,  travels through several connected pipes, each with a different pipe size (in bit rate) and expected  damage to the bits (in bit error rate). The bottleneck is typically in the wireless hops, nevertheless all the  constituents, and the protocols operating them, must be in full coordination and cooperation.

To this end, a paradigm shift is required in system design to host the semantic applications more efficiently over telecommunication networks. A list is provided in \tabref{Tab:paradigmchange} that summarizes  the pillars of such   changes in system design. The effect of this paradigm shift is more concretely summarized in the following, by describing the essential evolutions that a  communication system could highly benefit from to realize semantic communications.

\subsubsection{Radio Access Enhancements}

Current radio access technologies are designed with error-free communication mindset. Packets are either correctly decoded and delivered by the network, or dropped if the combination of \gls{fec} and automatic repeat request (ARQ) based techniques fail to guarantee the integrity of the packet---even a single bit of error in a packet will fail the whole. In contrast, lossy \gls{jscc} systems can  benefit from erroneous packets, especially if the rate of bit errors is anticipated. Enabling passage of damaged packets from radio access to the higher layers and through to the application is  therefore an essential element to enable lossy \gls{jscc}. 
\paragraph*{Additional considerations} Radio access protocols can benefit from new design paradigms in their use of \gls{fec}, \gls{arq} and radio resource management when moving from \gls{bler} based performance indicators (packet reliability over time) to \gls{ber} based quality of service (bit reliability within each packet). The proposed rate-adaptive and stable link design in \secref{sec:ratelessradiolink} is realized with uncoded modulated transmission, but in presence of time and frequency diversity such radio link can benefit from \gls{fec} and \gls{harq} based physical layer solutions. Stabilizing the radio link \gls{ber} with $1-\epsilon$ certainty can then be further optimized with the degrees of flexibility offered by channel coding and retransmission techniques. For example, \gls{fec} can be used to reduce the bit flipping ratio from the uncoded and modulated transmission, while \gls{harq} makes this process more efficient by leveraging time-diversity. Overhead  \gls{crc} bits which are normally used to check the decoding success and integrity of a packet, can now be used to estimate the bit flipping ratio within a packet. Hybrid \gls{arq} techniques which are typically used to increase chance of decoding success and improve throughput, can now be used to curb $\epsilon$.


\begin{table*}
\caption{Pillars of system design paradigm change from conventional communications to semantic/effectiveness  communications systems.}
\label{Tab:paradigmchange}
\footnotesize
\begin{tabularx}{\textwidth}{|>{\raggedright\arraybackslash}p{.2\textwidth}||>{\raggedright\arraybackslash}X|>{\raggedright\arraybackslash}X|}
\hline 
\cellcolor{lightgray}\textbf{System attribute} & \cellcolor{lightgray}\textbf{Conventional communication} & \cellcolor{lightgray}\textbf{Semantic \& Effective communication}  \\
\hline \hline
\cellcolor{verylightblue}\textbf{Transmission structure}     & Packet-ized transmission: \newline   Application exchanges  information bits in \emph{packets} with the network.   &  Streamlined transmission (bit pipe): \newline Communication is organized around \emph{frames}, which serve as temporal or spatial segments of application's  data.    
 \\ \hline
\cellcolor{verylightblue}\textbf{End-to-end link model}    & 
\xmark \quad Bit errors \newline
\xmark \quad Bit erasures \newline
\cmark \quad Packet erasures \newline
A packet is either delivered correctly or not delivered at all. The effect is that erasures could happen to the packets, but delivered packets are deemed \emph{error-free}.  &  
\cmark \quad Bit errors \newline
\cmark \quad Bit erasures \newline
\xmark \quad Packet erasures \newline
Erasure at the frame level is not acceptable, but damaged and missing bits are tolerated. Frames don’t need to be treated as indivisible units---parts of the frame may arrive intact, while others may be damaged or missing.\\  \hline 
\cellcolor{verylightblue}\textbf{Performance indication principle }    & Rate of successful delivery of packets over time, i.e., \gls{per} or \gls{bler}. &  Size of the \emph{delivered} frame (in bits, and as ratio of the original frame size) and the rate of damaged bits, i.e., \gls{ber} within each frame.    \\ \hline
\cellcolor{verylightblue}\textbf{Data integrity}      & Data integrity is enforced at the packet level.      & Data integrity is not strictly enforced at the frame level.    \\ \hline
\end{tabularx}
\end{table*}

\subsubsection{Enriched Interfacing and \gls{qos} monitoring} \label{sec:enrichedinterface}
\gls{qos} metrics in telecommunication networks are used to ensure compliance with the expectations of service defined in \gls{sla} that is between the application and the network. Packet reliability, latency and throughput (rate) are among the common \gls{qos} metrics defined for radio access systems. For semantic and effective communications, it is evident that new \gls{qos} metrics are required. Notably, the bit-pipe analogy of network's role suggests use of the following as new performance indicator metrics in semantic communication systems:
\begin{itemize}
    \item Bit flipping ratio $q$, measures the ratio of erroneous bits within a packet delivered to the application destination.
    \item Link stability $1 - \epsilon$, is a  measure of the stability of bit flipping ratio over time which is used to satisfy $\mathsf{Pr}[q > q_o] \leq \epsilon$.
    \item Maximum allowed puncturing size, $\bar{L}$, to limit the number of bits to be punctured out of a packet (bit erasures) for a rate-adaptive and stable link operation.
    \item Limit on the average puncturing length over time, $\mathsf{L}$, to incentivize the networks   to minimize puncturing attempts, i.e., $\mathbb{E} \left(  L\right) \leq \mathsf{L}$.
\end{itemize}
\paragraph*{Additional considerations} In the context of packetized communication, slight modification to the existing wireless communication systems together with enriched interfacing between the application and the network can be beneficial to the semantic applications too. Specifically, enabling multiple levels of \gls{qos} within the same packet   (as opposed to the single \gls{qos} level treatment of packets in current systems) can be beneficial. The application can flexibly choose to have multiple levels of \gls{qos} within the same packet, each level assigned to a different segment of  the packet. The network treats each segment according to the requested \gls{qos} level by the application, allowing the application to decide on the importance and significance of different segments of the packet in a dynamic way. This requires an enriched interfacing between the application and the network, enabling \emph{application-network symbiosis}. This allows time-varying segmenting and \gls{qos} level assignment to the segments by the application to be followed swiftly by the network.  

\subsubsection{Semantic Networking} 
Bit-pipe analogy for  the end-to-end  communication link requires all the layers of the stack to allow the passage of otherwise erroneous packets. While this starts in radio access protocols, particularly in physical layer, it extends to higher layers of the stack, including transport layer protocols. Most notably,  data integrity in \gls{tcp} is used to  guarantee that data is delivered accurately and without corruption, which hinders passage of erroneous packets through the network. Adjustments to those protocols are needed to securely relax those constraint, similar to \gls{udplite}, which allows partially damaged data to be delivered to the application \cite{larzon1999lighter}.
In case of an end-to-end link which has to go through multiple  hops  (see examples in \figref{fig:e2e}), it is important to allow seemless passage of the bits through every single hop without blocking the passage of damaged packets. The bottleneck hop, i.e., the hop that has the lowest capacity and/or causes the highest bit flipping ratio, determines the end-to-end experience. A rateless \gls{jscc} code such as the one proposed in this paper can smoothly make the best use  of such setting, by allowing  necessary puncturing of the bits in each hop. This significantly simplifies cross-layer protocol design approaches.
In case of satellite communications (the examples at the bottom of \figref{fig:e2e}), the large propagation delay between the ground and orbit satellites is not friendly to retransmission techniques such as \gls{arq}. Therefore, for semantic applications with limited latency budget, a one shot best effort transmission over the satellite link is most favorable.  In such case, rateless \gls{jscc} poses an additional benefit, that is, allowing to maximize the utilization of the one shot transmission by enabling the network to puncture the necessary number of bits out of the packet according to the estimated channel, and moving on to the next packet without the need for awaiting feedbacking and retransmission attempts.

\subsection{Open Research Problems}
\label{sec:openresearch}


Let us now discuss the open research problems and development challenges that need to be addressed 

\subsubsection{Resource Allocation and Scheduling}
Coexistence of  semantic applications with applications that use conventional communication links, over the same network platform, is inevitable. Given the  disparity in treatment of information bits between the two links (see \tabref{Tab:paradigmchange})  and the differences in \gls{qos} metrics, scheduling between the two types in multi-user settings becomes an open problem which requires new optimization and design frameworks.
\subsubsection{Proper Benchmarking}
The networking systems for multimedia communication are extremely intricate. They allow for handling a large multimedia packet in segments, where segments are from hierarchical source compression with unequal importance. The network treats  different segments with different \gls{qos} levels relative to their importance---but the network may be unable to trace back those segments to the same source and may be unaware of their potential overlap. As a result, the end to end link from the point of view of the application appears as a segment-wise erasure channel--that is, segments of a larger packet are either correctly delivered or lost in communication, while the erasure rate across segments are non-uniform. This could potentially create large overhead and repetition orders. Nevertheless, existing approaches should be considered as  the baseline for comparison with any new solution. Tractable mathematical modeling of those baseline methods can become very handy and are open for research.  


\subsubsection{Data Integrity for Streamlined Bit Pipes}
As outlined in \tabref{Tab:paradigmchange}, moving from  packet-ized transmission to streamlined ``bit pipe'' transmission appears to bring non-negligible gain for semantic communications. Providing efficient data integrity solutions in such cases requires further investigation, especially in the context of widely used internet protocols.

\subsubsection{Application and Network Symbiosis}
Admittedly, one of the main benefits of using \glspl{dnn} in source compression and \gls{jscc} solutions is their ability  to learn the distribution of the source and adapt the compression accordingly towards a better rate-distortion tradeoff. In turn, this requires well-tailored solutions to specific data formats and distributions. Such techniques may allow richer contextual information to be stored on devices, thus further reducing the compression size---this can be based on generic understanding of modes and contexts, e.g., using generative models or, can be distribution and data-set specific. This is an active research area which tightly relates to semantic communication. An interesting direction is to investigate solutions that account for the practical constraints of the networking systems and seek techniques that fit well with the telecommunication networks. Interfacing between the network and application can be more helpful when the exchange of \emph{side-information} works both ways and when it coveys enriched information about the state of the two sides, as already discussed in works like \cite{10201232}. Such interfacing could benefit semantic communication systems too (see \secref{sec:enrichedinterface}) which requires more research and development effort.   

\subsubsection{Physical Layer Research}

As was discussed in \secref{sec:discussionofresults}, the current coding, modulation schemes, and retransmission techniques utilized in lower and upper parts of the 5G new radio physical layer, are mainly tailored  towards error-free packet-ized communications (see \tabref{Tab:paradigmchange}). For the semantic applications to benefit from a more flexible radio link with streamlined transmission, modulation shaping using machine learning, coded modulation techniques, and retransmission mechanisms that target \gls{ber}  stabilization  can be beneficial. \Gls{ai}-native air interface and \gls{dnn} based physical layer solutions such as \cite{honkala2021deeprx} could provide a fast track towards this end, but further research and development is needed  to generalize those to different applications and network state distributions. 

\subsubsection{Rate-Distortion-Memory Tradeoffs} 
In the context of deep learning based \gls{jscc}, the matter of caching of the trained neural networks is an important practical consideration. The trained neural networks must be stored (and in many cases, communicated before storage) to be useful for the application, while the cost of such caching is non-negligible. A relatively small neural network with 10 million trainable parameters stored with 32-bit floating point representations requires approximately 38 megabytes of storage space. As the neural network size grows and as the dataset becomes more specific, the relative cost of such storage increases. Therefore, one must study the matter of generalizability of the trained DNNs and the impact of  it on  JSCC performance. The alternative direction is to build very small neural networks that are tailored to  each specific dataset. Neural network compression methods \cite{neill2020overview} could also be exercised, but the impact of such  compression on JSCC performance remains an open problem. In fact, one may see the trained neural networks as the shared \emph{context} between the source and destination points for the semantic communication problem and study the tradeoff between context memory size, semantic distortion and communication rate.

\subsubsection{Other Open Problems from the Literature}
Several works in the literature have nicely reviewed the state-of-the-art in semantic communications and outlined the open research problems.  \cite{gunduz2022beyond} provides a comprehensive overview of semantic and task-oriented communications, focusing on integrating semantics and goals into communication system design. The authors discuss the limitations of traditional communication systems that focus solely on reliable bit transmission, and highlight the potential of deep learning-based \gls{jscc} techniques for efficient video delivery. Interoperability and hardware design for those techniques require further investigation. 
Specific data types require specific treatement when it comes to semantic compression and semantic JSCC. An intereting example is studied in  \cite{liang2024semantic} for synchronized sound communication, which demonstrates an interesting inter-working between synchronization and semantic communication. 

Use of information theoretic tools for formalization and quantification of semantics and effectiveness of information are required, as was also pointed out by \cite{qin2021semantic}.  
It is not concretely accepted whether measures such as semantic entropy and semantic level rate-distortion theory are the correct metrics for the job. \cite{luo2022semantic} also points out that there is need for more theoretical work on the topic, especially in conjunction with deep learning based codes. Additionally, the capacity of semantic networking is  an open problem \cite{shi2021semantic} and a comprehensive mathematical framework to evaluate the performance limits of a semantic communications system requires further research effort.

Network-level \gls{qoe} quantification for semantic applications contrasts with conventional communication, as pointed out in \cite{shi2021semantic}, and needs a rethinking. In \secref{sec:enrichedinterface} we discussed this from \gls{qos} perspective and proposed a set of new \gls{qos} metrics suitable for semantic communication. 
In an interesting take, \cite{you2024next} addresses the topic from communication hardware  perspective and discusses the next generation advanced transceivers that are friendlier with semantic communication, and raises the relevant open research problems in that context, with interesting connection to deep learning based solutions.

Data integrity (see \tabref{Tab:paradigmchange}), data security, and privacy are other aspects of semantic communications that needs rethinking. \cite{getu2024survey} argues that \gls{6g} will be highly tied with semantic communication and sees security and data privacy as one of the main challenges that need to be addressed by future research.

\section{Summary and Conclusions}
\label{sec:conclusions}

    The main thesis of this work is as follows: semantic communication is mostly a source coding problem, and less of a communication networking problem. To tackle errors caused over the communication link, the source coding should be error-tolerant, or, what we refer to as JSCC in this paper, which is deemed to outperform the separate design counterpart. The performance promised by \gls{jscc}  for semantic communication can be derived without the need to re-design the existing networks from scratch. With some essential modification to the existing systems---which are built with technical communication paradigm in mind---networks can turn the underlying channel(s) into a suitable end-to-end link with controlled uncertainty and allow the applications to run semantic communications, e.g.,  with deep-learning based \gls{jscc}.

    To this end, we  introduced  rateless JSCC, designed and optimized for a continuum of coding rates, which enables the application to efficiently  JSCC encode the source signal without having knowledge of the channel state. To complement this,  rate-adaptive and stable communication link operation was proposed which allows adapting the rate of the already encoded codeword to the channel capacity by  puncturing bits out of it. The network  maintains the bit flipping ratio across time frames. Together, the rateless JSCC in the application, and the rate-adaptive and stable link in the network, form a cooperative joint venture to efficiently adapt the JSCC code rate to the channel state, without exchange of \gls{csi}. 
    
    We demonstrated the feasibility of this joint venture using autoencoder based  JSCCs. Specifically, a new family of  autoencoder rateless JSCC codes were introduced and tested for reconstruction loss of image signals and demonstrated powerful performance and resilience to variation of channel quality.
   
    Next, we put things in perspective, and outlined the essential modifications needed for the next generation communication networks to enable such rateless JSCC and other semantic communication solutions. We studied the practical concerns regarding semantic communication  and provided a blueprint for networking system design, especially towards the anticipated 6G systems.

\bibliographystyle{IEEEtran}
\bibliography{references}

\vfill\pagebreak

\end{document}